\documentclass[amsmath,amssymb,prb,twocolumn]{revtex4-2}

\usepackage[utf8]{inputenc}
\usepackage{graphicx}
\usepackage{dcolumn}
\usepackage{relsize}
\usepackage{bm}
\usepackage{appendix}
\usepackage{braket}
\usepackage{color}
\usepackage{hyperref}

\begin{document}

\title{Solution of the mean-field Hubbard model of graphene rectangulenes}

\author{Amador García-Fuente$^{1,2}$}
\author{Jaime Ferrer$^{1,2}$}
\email{ferrer@uniovi.es}
\affiliation{$^1$ Departamento de Física,  Universidad de Oviedo,  33007 Oviedo, Spain}
\affiliation{$^2$ Centro de Investigación en Nanomateriales y Nanotecnología, Universidad de Oviedo-Consejo Superior de
Investigaciones Científicas, 33940 El Entrego, Spain}

\begin{abstract}
We present the analytical solution of the mean-field Hubbard model of undoped and doped 
graphene rectangulenes. These are non-chiral graphene rectangles of arbitrary length and width, and 
nanoribbons are thus narrow rectangulenes. We rewrite the Hubbard
model in the basis of exact bulk and edge non-interacting eigen-states, and provide analytical expressions for 
the Coulomb integrals. We also present a general mean-field decoupling of the Hamiltonian, that is valid at
any temperature and doping. We introduce the edge-only doping regime and discuss 
the paramagnetic, ferromagnetic and antiferromagnetic mean-field solutions in this regime. We calculate 
explicitly at zero temperature the 
eigen-energies, occupations, spin densities and addition
energies of rectangulenes with lengths and widths ranging from a nanometer to several hundreds of them.
We rewrite the exact mean-field tight-binding Hamiltonian back in the site-occupation basis.
\end{abstract}

\maketitle

\vspace{0.1cm}

\section{Introduction}
\label{Section:Introduction}
The experimental demonstration of the ability to isolate single graphene sheets \cite{Novoselov2004} promoted the
vision that atomic-scale two-dimensional nanoelectronics and nano-optics could be a viable future
technology \cite{Westervelt2008}. Elementary units of the graphene lego would then range from
simple graphenoid molecules such as acenes \cite{Yang2016,Eisenhut2020}, graphene nanoribbons (GNR) and all the
way up to graphene flakes. We introduce here graphene rectangulenes. These are graphene flakes of rectangular 
shape having two zigzag and two armchair edges. Achiral finite-length GNRs are therefore narrow rectangulenes.
Part of the interest in graphene nanostructures stems from the old prediction by Dresselhaus and coworkers 
that GNRs having zigzag terminations could host edge  states\cite{Dresselhaus1996}. GNRs' peculiar electronic 
and magnetic structure were the subject of intense theoretical work for the first years after the discovery of graphene 
\cite{Brey2006,Wakabayashi2010,Guinea2009,Akhmerov2011,Son2006,Yang2007,Rossier2008,Jung2009a,Jung2009b}. 

Parallel efforts to fabricate GNRs by unzipping carbon nanotubes were only partially successful because the graphene
edges were quite defective \cite{kosynkin2009}. However, GNRs having atomically-precise edges were finally synthesized 
by bottom-up techniques \cite{Cai2010}. This breakthrough opened the door to a plethora of subsequent developments in
GNR fabrication and characterization \cite{Kimouche2015,Wang2016,Talirz2017,Lawrence2022,Oteyza2023}. 
The topological nature of edge states \cite{Ryu2002,Delplace2011} and the connection between GNRs and the 
Schrieffer-Heeger-Su 
(SSH) model  \cite{Su1979,Laszlobook} were also uncovered and analyzed both theoretically \cite{Cao2017} and
experimentally \cite{Groning2018,Rizzo2018}. GNR edge states were predicted to be magnetic \cite{Fujita96,Lee2005,Son2006},
so that magnetism at the edge was paid attention throughout these years \cite{Slota2018,Lawrence2020}.
GNRs were also explored  in optics for their potential utility as plasmon waveguides \cite{Christensen2012,Fei2015}.
Recently, single armchair GNRs of precise width have been deposited onto ultra-clean graphene gaps, therefore 
creating all-carbon single electron transistors displaying quantum dot behavior \cite{Niu2023,Zhang2023a,Zhang2023b}.
Both graphene nanogaps and atomically-precise ribbons tend to be as long as several tens of 
nanometers at least, thus containing from thousands to hundreds of thousands of carbon atoms.

GNRs have been simulated both by Density Functional Theory (DFT) \cite{Son2006,Yang2007}
and  by the mean-field (MF) Hubbard model \cite{Wakabayashi98,Rossier2008}. These two types of numerical
simulations estimate successfully many of the electronic, magnetic and optical properties of GNRs, but also 
have several shortcomings. First, these simulations are numerically costly for lengths of about 20-30 nm, while
simulations for lengths longer than 100-200 nm are beyond the power of today's computers. Second, these 
simulations fail to estimate correctly the addition energies of correlated electron 
nano-structures \cite{Carrascal12,Carrascal15} such as for example those of graphene nanoribbons.
Third, numerical simulations are akin to numerical experiments, and therefore must be complemented 
and guided by analytical results in order to understand the underlying physics in full depth. 

One-body tight-binding models are easily solved analytically for infinite lattices, where periodic boundary 
conditions can be applied. Similarly, metallic nanostructures such as disordered quantum dots can generally be 
addressed by Random Matrix Theory up to the Thouless energy scale \cite{Kurland00}. 
Doped rectangulenes are highly ordered metallic nano-structures whose boundaries carry functionality in the form 
of spin-active edge states, so boundary conditions do matter for them even when they approach bulk sizes 
\cite{Carrascal12b}.
Finite-size tight-binding lattices can in principle be solved by applying suitable boundary conditions 
that mix same-energy Bloch waves. This is easily done in one dimension, while difficulties arise in two and higher 
dimensions, because reflection at the boundaries mix too many (or infinite) waves, resulting frequently in chaotic 
cavities. We have recently been able to solve analytically the tight-binding model of non-chiral 
rectangulenes of arbitrary length and width by mapping the model to a wave-guide of $M_y$ finite-length SSH 
chains \cite{Amador2023}. This solution has allowed us to unveil explicitly the bulk-boundary correspondence 
\cite{Hasan2010} in graphene. We have also provided an accurate mapping between DFT-simulated $N=5,\,7\,9$ GNRs
and a simple two-site Hubbard model for the edge states of those ribbons. We are not aware of any 
previous solution of a one-body tight-binding model in a finite size non-trivial two-dimensional lattice.

Analytical solutions of 
interacting electron systems are extremely hard to find, and only a few simple models have been solved in the past hundred 
years. These include one-dimensional models that can be addressed via Bethe-ansatz \cite{Lieb68,Wiegmann80,Andrei80,carmelo92}, 
as well as the Schrieffer-Heeger-Su or the Falicov-Kimball models \cite{Schrieffer,Falicov}.
Furthermore, many-body tight-binding models in two-dimensions have only been solved numerically and only for very small 
sizes. Mean-field solutions enable researchers to reach much larger sizes, but these can still contain only 
a few hundred or within the DFT context a few thousand atoms. Beyond-Mean-field numerical simulations of finite lattices
are restricted to less than a hundred atoms at the moment.

We expand here our previous development \cite{Amador2023} to address the Hubbard model of graphene rectangulenes.
We perform a basis change from site creation and destruction operators to the basis of bulk and edge
eigen-states. We are thus able to write down analytical expressions for the Coulomb integrals. 
The resulting Hamiltonian is therefore fully known with explicit formulae in terms of $U/t$. 
The essence of our solution thus consists on the analytical decomposition of the two-body Hubbard Coulomb integral 
in terms of the eigen-states of the non-interacting graphene rectangle Hamiltonian.
We note that this kind of decoupling has not been explored yet for the Hubbard model of a graphene device. 
It has however an illustrious past. Indeed, decoupling the Coulomb interaction of the Jellium model in terms of eigen-states 
of the kinetic energy term enabled physicist to develop non-relativistic quantum field theory in the fifties and sixties
of the past century \cite{Ashcroft,Pines,Fetter} and was essential also in the establishment of Density Functional 
Theory \cite{Gell-Mann,Perdew}. 
We then go on, and write down the paramagnetic, ferromagnetic and antiferromagnetic mean field solutions of the
Hubbard model of a graphene rectangulene. This new development allows us to describe undoped and doped rectangulenes of 
any  given size, ranging from the smallest ribbons all the way to lengths and widths of several hundreds of nanometers or 
even micrometers, therefore extending the range of numerical
simulations of GNRs at least a thousandfold. This is to the best of our knowledge, the first analytical solution
of a many-body model at the Mean-Field level in a finite lattice or arbitrary size.
We demonstrate that our analytical MF solutions deliver all the physics that had to be 
previously computed numerically \cite{Son2006,Yang2007,Wakabayashi98,Rossier2008}. 
We illustrate this by computing the rectangulene band structures, that
agree well with the published results for infinite-length GNRs, as well as the mean-field addition energies, 
energy differences among the mean field phases, and the site-charge and site-spin occupations. These last
quantities allow us to return to the real space site picture, and rewrite the self-consistent mean field Hubbard model
in the tight-binding basis. The figures shown 
in this article have been plotted with the aid of a simple matlab script running in a laptop. Site occupation
calculations lasted the longest, with rectangulenes of size 200 nm$\times$ 400 nm taking four to five minutes.
Our analytical expressions for the eigen-energies, the eigen-functions and the Coulomb integrals 
can also be implemented in GW or other beyond-Mean-Field numerical approaches extending 
enormously the size of rectangulenes that can be simulated. This should allow theoreticians and simulators to make
direct contact with real-life electrical transport and excitonics experiments.

The layout of this article is as follows. Section \ref{Section:sshmapping} summarizes the key results of our previous 
solution on the non-interacting tight-binding model \cite{Amador2023}, and introduces the notation and 
terminology needed henceforth. Section \ref{Section:solution} explains the change from the original site 
creation and destruction operators to the basis of bulk and edge eigen-states, and determines all the 
Coulomb integrals. The final result in the section is the complete reformulation of the
Hubbard model in the eigen-state basis. Section \ref{Section:meanfieldsolutions} develops the generic
MF decomposition and then shows the PM, FM and AFM solutions. We compute eigen-energies, addition
energies and site charge and spin occupations. These occupations allow us to rewrite the MF Hamiltonian
in the site-basis. We introduce the {\it edge-only doping} regime and use it to analyse the impact of doping
the rectangulenes. We also discuss how to release this approximation to address larger doping regimes.
Section \ref{Section:conclusions} summarizes our results and closes this article.

\begin{figure}[ht]  \centering
  \includegraphics[width=\columnwidth]{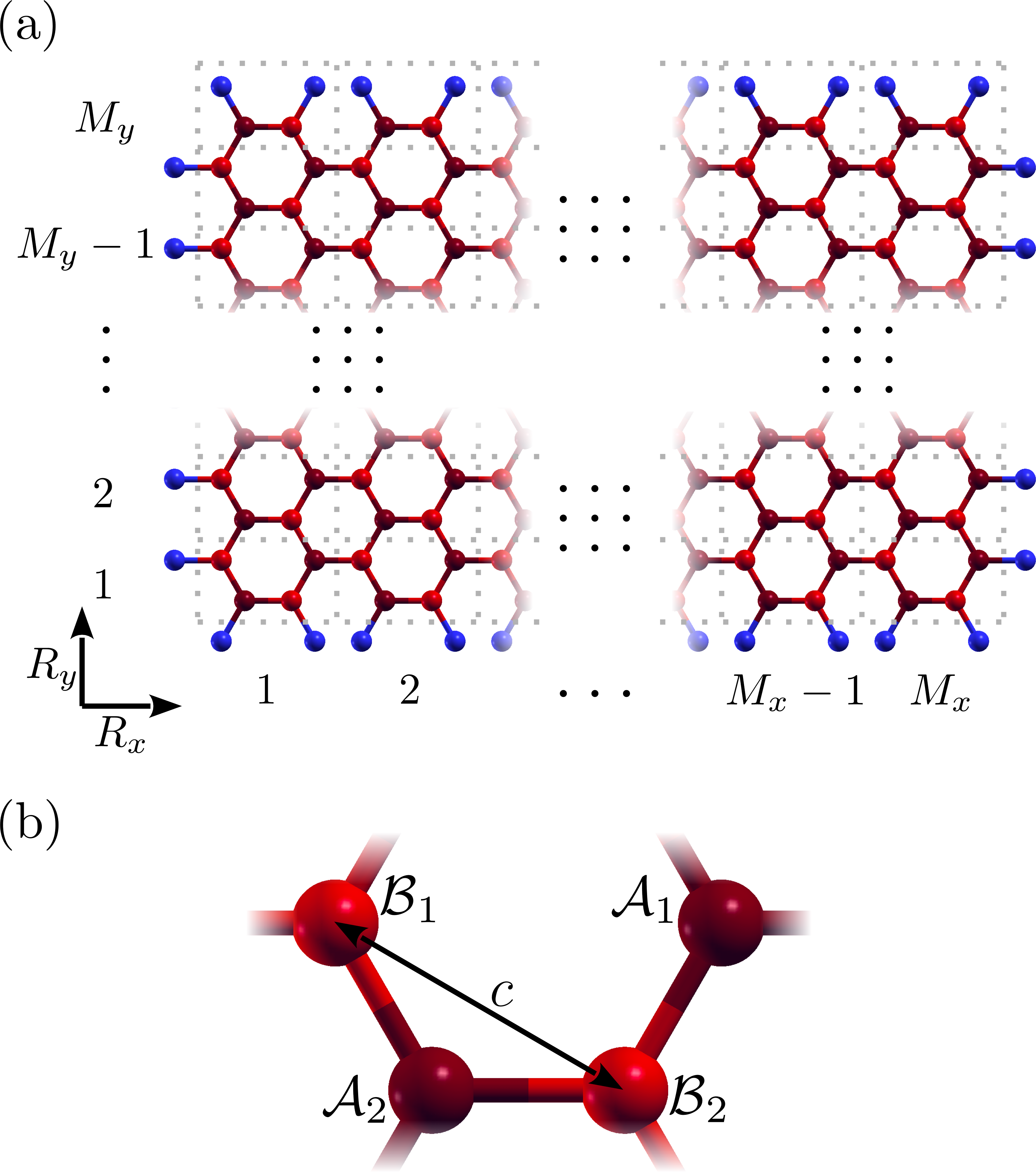}
  \caption{(a) Rectangulene with dimensions $M_x \times M_y$. ${\cal A}$/${\cal B}$ atoms are indicated by dark/bright red circles. Fake
  atoms, where wave-function coefficients are set to zero, are indicated by blue circles. Each unit cell is surrounded with 
  a grey dotted box. (b) Each unit cell contains two ${\cal A}$
  and two ${\cal B}$ atoms, whose internal coordinates are written in Eq. (\ref{eqn:internalcoordinates}).
  The black arrow indicates the lattice constant.}
  \label{Figure:geometry}
\end{figure}

\section{Solution of the tight-binding model of a graphene rectangulene}
\label{Section:sshmapping}
We summarize below key results of our solution of the tight-binding Hamiltonian \cite{Amador2023}

\begin{eqnarray}
\hat{H}^0=\sum_{{\bf R}i \sigma}\sum_{a=A,B}\,\epsilon_0\,\hat{n}^a_{{\bf R} i \sigma}
-t\,\sum_{< {\bf R} i \sigma, {\bf R'} i' \sigma' >}\left(\hat{a}^\dagger_{{\bf R} i\sigma} \hat{b}_{{\bf R'} i'\sigma'}+ c.c.\right)\nonumber\\
\end{eqnarray}
of the rectangulene 
drawn in figure (\ref{Figure:geometry}) (a). The figure shows that the rectangulene is pierced by small 
rectangles that constitute the different unit cells. Cell coordinates are ${\bf R}=(R_x,\,R_y)$, where $R_x$ and $R_y$ are 
integer numbers running from 1 to $M_x$ and from 1 to $M_y$, respectively. The width along the Y-axis can also be characterized 
by the number of horizontal bonds $N=2\,M_y-1$. The unit cell, depicted in Figure (\ref{Figure:geometry}) (b), contains 
two $\cal{A}$-atoms and two $\cal{B}$-atoms, that we label ${\cal A}_1$, ${\cal A}_2$, ${\cal B}_1$ and ${\cal B}_2$, respectively. 
We measure lengths along the cartesian X- and Y-axes in units of $\sqrt{3}\,c$ and $c$, respectively, 
where $c=2.46$ \AA\ is graphene's 
lattice constant. The atoms' coordinates in the re-scaled orthogonal reference frame are then 

\begin{eqnarray}
\label{eqn:internalcoordinates}
\begin{matrix}
{\bf r}_{{\cal A}_1} &=& \left(\begin{matrix}0\\0\end{matrix}\right),\,\,\,
{\bf r}_{{\cal B}_1} &=& \left(\begin{matrix}-2/3\\0\end{matrix}\right)\\
{\bf r}_{{\cal A}_2} &=& \left(\begin{matrix}-1/2\\-1/2\end{matrix}\right),\,\,\,
{\bf r}_{{\cal B}_2} &=& \left(\begin{matrix}-1/6\\-1/2\end{matrix}\right)
\end{matrix}
\end{eqnarray} 

\begin{figure}[ht]  \centering
  \includegraphics[width=\columnwidth]{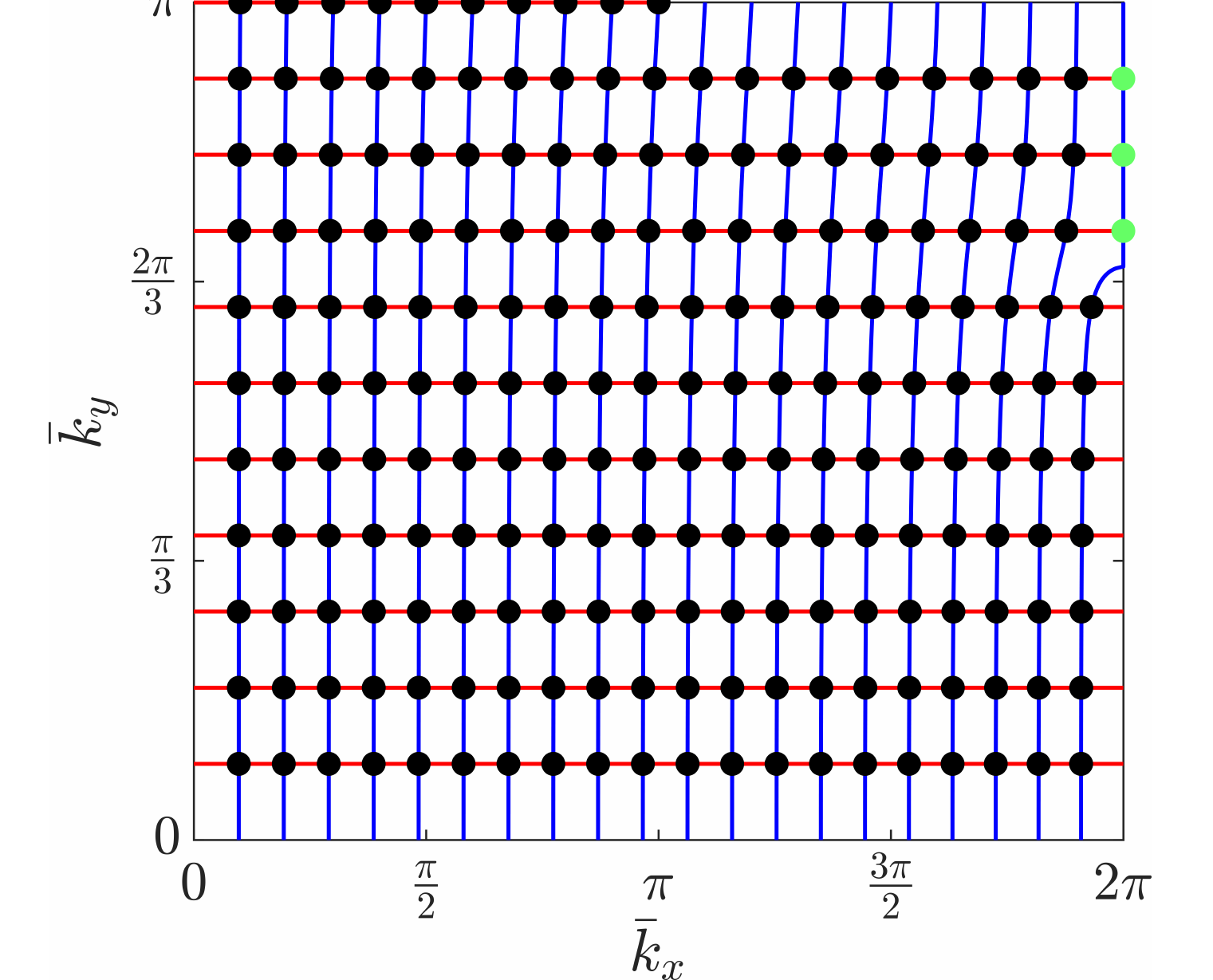}
  \caption{Two-dimensional plot of the mesh of allowed ${\bar{\bf k}}$-vectors for a rectangulene with dimensions
           $(\,M_x,\,M_y\,)\,=\,(\,10,\,11\,)$. The red lines correspond to the $\bar{k}_y$ quantized values.
           Blue lines correspond to solving Eq. (\ref{Equation:kx_quantization}) for $\bar{k}_y$ as a 
           function of $\bar{k}_x$. Black dots at the intersections between blue and red lines correspond
           to bulk states. Green dots correspond to edge states.  \label{Figure:kgrid}}
\end{figure}

The rectangulene boundary conditions determine the set of eigen-states of the rectangulene. These correspond to the 
allowed ${\bf \bar{k}}=(\bar{k}_x,\,\bar{k}_y)=(\sqrt{3}\,k_x\,c,\,k_y\,c)$ wave-vectors.  Notice that we have introduced here 
dimensionless units for consistency with our choice of real-space unit lengths, as well as because the algebraic 
expressions are simpler. Figure \ref{Figure:kgrid} summarizes the ${\bf k}$-vector grid, that covers a rectangular area 
in reciprocal space, where $\bar{k}_x\in (0,\,2\pi]$ and $\bar{k}_y\in (0,\,\pi]$ (note that $\bar{k}_x=0$ and  
$\bar{k}_y=0$ are excluded because $\sin(0)=0$). The segment
$\bar{k}_x\in(\pi,\,2\,\pi]$ with $\bar{k}_y=\pi$ is excluded to avoid state double-counting.  
The $\bar{k}_y$ quantization condition is seen in the figure as horizontal red lines at values given by 
\begin{eqnarray}
\bar{k}_y=\bar{k}_m=\pi\,\frac{m}{M_y},\,\,m=1,,2,,...,M_y
\end{eqnarray}
For a given $\bar{k}_m$, there exist a set of $2 M_x$ $\bar{k}_x$ wave-vectors that we denote $\bar{k}_{m\alpha}$ 
where $\alpha=1,\,2,\,...,\,2 M_x$. These 
occur at the intersections of the red and blue lines in Figure \ref{Figure:kgrid}. The resulting black and green
dots in the figure mark the allowed $(\bar{k}_x,\,\bar{k}_y)=(\bar{k}_{m\alpha},\,\bar{k}_m)$ that define the 
eigen-states of the rectangulene. We find a critical $y$-wavevector
\begin{eqnarray}
\bar{k}_m^c=2\,\cos^{-1}{\frac{M_x}{2\,M_x+1}}\gtrsim \frac{2}{3}\,\pi
\label{Equation:kmc}
\end{eqnarray}
so that all $\bar{k}_m$ smaller or larger than $\bar{k}_m^c$ have either $2\,M_x$ or $2\,M_x-1$ $\bar{k}_{m\alpha}$ 
real wave-vectors, respectively. These corresponds to bulk rectangulene states, and are marked by black dots in 
Figure \ref{Figure:kgrid}. The missing $\bar{k}_{m\alpha=2\,M_x}$ wave-vector whenever $\bar{k}_m>\bar{k}_m^c$ is sketched as a 
green dot in the figure and is found by letting $\bar{k}_{m\alpha=2\,M_x}=2\,\pi-i\,q_m$ become complex.
The corresponding eigen-state is an edge state at the zigzag edges whose decay length is $q_m^{-1}$.
As a consequence, the number of allowed ${\bf k}$-vectors is equal to the number of unit cells in the rectangulene,
$2\,M_x\times (M_y-1/2)=M_x\times N $, and the number of edge states is 
\begin{eqnarray}
N^\mathrm{edge}=\mathrm{Floor}\left(\left(1-\frac{2}{\pi}\cos^{-1}{\frac{M_x}{2\,M_x+1}}\right)\,M_y\right)
\label{Equation:nedge}
\end{eqnarray}
where the mathematical Floor$(x)$ function takes as input any real number $x$, and gives as output the greatest 
integer less than or equal to $x$. 
The explicit values of the bulk $\bar{k}_{m\alpha}$ wave-vectors are found by replacing $\bar{k}_y$ 
by $\bar{k}_m$ in the equations for graphene's  order parameter, Bloch Hamiltonian component and Bloch phase:
\begin{eqnarray}
\Delta_y&=&2\,\cos{(\bar{k}_y/2)}\\
f_{x y}&=&f^R_{x y}+i\,f^I_{x y}=1+\Delta_y\,e^{i\,\frac{\bar{k}_x}{2}}\nonumber\\
\tan\theta_{x y}&=&\frac{f^I}{f^R}=\frac{\Delta_y\,\sin{(\bar{k}_x/2)}}{1+\Delta_y\,\cos{(\bar{k}_x/2)}}\nonumber
\end{eqnarray}
and solving for $\bar{k}_x$ the equation
\begin{eqnarray}
M_x\,\bar{k}_x+\theta_{xy}= \alpha \pi
\label{Equation:kx_quantization}
\end{eqnarray}
where $\bar{k}_x\in(0,\,2\pi]$ and $\alpha$ is an integer number.
Alternatively, Equation (\ref{Equation:kx_quantization}) can be seen as an implicit equation for $\bar{k}_y$ as a function
of $\bar{k}_x$, that we plot as blue lines in Figure \ref{Figure:kgrid}. Then, the grid of allowed 
$(\bar{k}_{m\alpha},\,\bar{k}_m)$ is given by the all the intersections of the red and blue lines.
The remaining edge wave-vector is the solution of the equation
\begin{eqnarray}
\tanh{(M_x\,q_m)}=\frac{\Delta_m\,\sinh{(q_m/2)}}{1-\Delta_m\,\cosh{(q_m/2)}}
\label{Equation:qedge}
\end{eqnarray}
where $\Delta_m=2\,\cos{(\bar{k}_m/2)}$.

The rectangulene bulk and edge eigen-energies are found by inserting the grid of allowed wave-vectors into
graphene's bulk dispersion relation. We find
\begin{eqnarray}
\label{Equation:eigenenergies}
\epsilon_{m\alpha\tau}^B&=&\tau\, \epsilon_{m \alpha}^B=\tau\,\sqrt{1+\Delta_m^2+2\,\Delta_m\,
\cos{(\bar{k}_{m\alpha}/2)}}\\
\epsilon_{m\tau}^E&=&\tau\,\epsilon_m^E=\tau\,\sqrt{1+\Delta_m^2-2\,\Delta_m\,\cosh{(q_m/2)}}\nonumber
\end{eqnarray}
The band index $\tau =\pm 1$ labels the two eigen-states existing for each wave-vector, so that the number of eigen-states
is equal to the number of atoms $2\,M_x\,N$. 
The second-quantized version of the diagonalized hamiltonian is 
\begin{eqnarray}
\hat{H}^0&=& \sum_{m \alpha \tau \sigma} \,\tau\,\epsilon^B_{m\alpha}\,\hat{n}^B_{m\alpha\tau\sigma}+
\sum_{m\tau\sigma} \,\tau\,\epsilon^E_{m}\,\hat{n}^E_{m\tau\sigma}
\end{eqnarray}
where $\hat{n}^B_{m\alpha\tau\sigma},\,\hat{n}^E_{m\tau\sigma}$ are the number operators for the bulk and edge eigen-states.
The explicit expressions for these eigen-states are
\begin{eqnarray}
\label{Equation:eigenstates}
\ket{\phi_{m \alpha \tau}}&=&\,\sum_{R_{x,y}=1}^{M_{x,y}}\,\sum_{i=1,2}\,
\frac{2\, f_{m,i}(R_y)}{{({\cal M}_x\,M_y\,\Lambda^\phi_{m\alpha})^{1/2}}}\,
\ket{{\bf R}^i}\,\bra{\phi_{m\alpha,i}(R_x)}\nonumber\\\\
\ket{\psi_{m \tau}}&=&\,\sum_{R_{x,y}=1}^{M_{x,y}}\,\sum_{i=1,2}\,
\frac{2\, f_{m,i}(R_y)}{({\cal M}_x\,M_y\,\Lambda^\psi_m)^{1/2}}\,\ket{{\bf R}^i}\,\bra{\psi_{m,i}(R_x)}\nonumber
\end{eqnarray}
with 
\begin{eqnarray}
\label{Equation:ffunctions}
{\cal M}_x&=& 4 M_x+1\\
f_{m,i}(R_y)&=&\sin{(\bar{k}_m\,(R_y-d_i))}\nonumber\\
\ket{{\bf R_i}}&=&\left(\ket{{\bf R}, {\cal A}_i},\,\ket{{\bf R}, {\cal B}_i}\right),\,\,i=1,\,2\nonumber\\
\bra{\phi_{m\alpha,i}(R_x)}&=&\left(\begin{matrix}\phi_{m\alpha,i}^{\cal A}\\\phi_{m\alpha,i}^{\cal B}\end{matrix}\right)\nonumber\\
&=&\left(\begin{matrix}-\tau\,(-1)^{\alpha} \sin{(\bar{k}_{m\alpha}\,(R_x-d_i))}\\\sin{(\bar{k}_{m\alpha}\,(M_x+1-(R_x+d_i)))} \end{matrix}\right)\nonumber\\
\bra{\psi_{m,i}(R_x)}&=&\left(\begin{matrix}\psi_{m,i}^{\cal A}\\\psi_{m,i}^{\cal B}\end{matrix}\right)\nonumber\\
&=&(-1)^{2d_i} \left(\begin{matrix}-\tau\,\sinh{(q_m\,(R_x-d_i))}\\\sinh{(q_m\,(M_x+1-(R_x+d_i)))} \end{matrix}\right)\nonumber
\end{eqnarray}
where $d_1=0$ and $d_2=1/2$, and the normalization factors are 
\begin{eqnarray}
\label{Equation:normalization}
\Lambda^\phi_{m \alpha}&=&F^1_{m\alpha}-\delta_{\bar{k}_m,\pi}\,F^2_{m\alpha}/{\cal M}_x\\
\Lambda^\psi_m&=&G^1_m\,\sinh{({\cal M}_x\,q_m/2)}\nonumber
\end{eqnarray}
We have introduced here the following functions
\begin{eqnarray}
F^1(k)&=&1-\frac{\sin{({\cal M}_x\,k/2)}}{{\cal M}_x\,\sin({k/2})}\\
F^2(k)&=&1-\frac{\cos{({\cal M}_x\,k/2)}}{{\cal M}_x\,\cos{(k/2)}}\nonumber\\
G^1(q)&=&\frac{1}{{\cal M}_x\,\sinh{(q/2)}}-\frac{1}{\sinh{({\cal M}_x\,q/2)}}\nonumber\\
G^2(q)&=&\frac{1}{{\cal M}_x\,\cosh{(q/2)}}-\frac{1}{\cosh{({\cal M}_x\,q/2)}}\nonumber
\end{eqnarray}

\noindent
where $F^1_{m\alpha} = F^1(\bar{k}_{m\alpha})$, $G^1_{m} = G^1(q_{m})$ and so forth.

We close this section by noting that the solution outlined above can also be understood by realizing that we have 
decomposed the rectangulene as
a wave-guide of $M_y$ open-ended SSH chains having $2\,M_x$ sites each and topological order parameter $\Delta_m$. 
Those chains having $\Delta_m<1$ or $\Delta_m>1$ are topological or trivial because their winding number is 1 or 0, 
respectively \cite{Laszlobook}.
The bulk-boundary correspondence \cite{Hasan2010} is explicitly established by noting that the winding number condition 
enters into the $\bar{k}_{m\alpha}$ and $q_m$ quantization equations (\ref{Equation:kx_quantization}) and (\ref{Equation:qedge}).

\section{Hubbard model of a graphene rectangulene}
\label{Section:solution}

\subsection{Conventional Mean-Field treatments of the Hubbard model}
The rectangulene's Hubbard Hamiltonian is
\begin{eqnarray}
\label{eqn:Hubbard}
\hat{H}&=&\hat{H}^0+\hat{V}^{ee}=\hat{H}^0+ U\,\sum_{{\bf R} i}\,\hat{n}_{{\bf R} i \uparrow} \hat{n}_{{\bf R} i \downarrow}
\end{eqnarray}
Notice that we shall address in this manuscript only the positive-$U$ Hubbard model as is adequate for graphene rectangulenes.

\subsection{Transformation to the eigen-state basis}
\label{Subsection:basis}
The approach proposed in this manuscript is initiated by expanding site-creation and annihilation operators 
in the basis of rectangulene eigen-states
\begin{eqnarray}
\left(\begin{matrix}\hat{a}_{{\bf R} i \sigma}\\\hat{b}_{{\bf R} i \sigma}\end{matrix}\right)=&&
\sum_{m \alpha \tau} 
\left(\begin{matrix}\braket{{\bf R},{\cal A}_i\,|\,\phi_{m \alpha \tau}}\\\braket{{\bf R},{\cal B}_i\,|\,\phi_{m \alpha \tau}}\end{matrix}\right)\,\hat{\phi}_{m \alpha \tau \sigma}+
\\&+&\sum_{m \tau}\left(\begin{matrix}\braket{{\bf R}, {\cal A}_i\,|\,\psi_{m \tau}}\\\braket{{\bf R}, {\cal B}_i\,|\,\psi_{m \tau}}\end{matrix}\right)
\,\hat{\psi}_{m \tau \sigma}\nonumber
\end{eqnarray}
The site number operators can be expressed in terms of bulk and edge pieces as follows
\begin{eqnarray}
\label{Equation:sitenumber}
\hat{n}_{{\bf R}i\sigma}&\approx&\hat{n}^B_{{\bf R}i\sigma}+\hat{n}^E_{{\bf R}i\sigma}\\
\hat{n}^B_{{\bf R}i\sigma}&\approx&\,\sum_{m \alpha}\,\frac{4\,f_{m,i}^2}{{\cal M}_x\,M_y\,\Lambda^\phi_{m\alpha}}\,
\left(\begin{matrix}(\phi_{m\alpha,i}^{\cal A})^2\\(\phi_{m\alpha,i}^{\cal B})^2\end{matrix}\right)\,
\hat{n}^B_{m\alpha\sigma}\nonumber\\
\hat{n}^E_{{\bf R}i\sigma}&\approx&\sum_{m}\, \frac{4\,f_{m,i}^2}{{\cal M}_x\,M_y\,\Lambda^\psi_m}\,\,
\left(\begin{matrix}(\psi_{m,i}^{\cal A})^2\,(\hat{n}^E_{m\sigma}-\hat{P}_{m\sigma})\\(\psi_{m,i}^{\cal B})^2
(\hat{n}^E_{m\sigma}+\hat{P}_{m\sigma})
\end{matrix}\right)\nonumber
\end{eqnarray}
Notice that the above equations are approximate because we have dropped crossed bulk-edge, as well as 
bulk $(m,\,\alpha) \leftrightarrow (m',\,\alpha')$, bulk $(m,\,\alpha,\,\tau)\leftrightarrow (m,\,\alpha,\,\tau')$ 
and edge $m\leftrightarrow m'$ terms. We have however kept inter-band edge states 
$(m,\,\tau)\leftrightarrow (m,\,\tau')$ because these are degenerate, meaning that perturbations 
entangle them. This poor man's approximation is a common strategy in many-body treatments; here, it is justified 
at the outset because the eigen-energies computed below agree extremely well with known numerical results for 
infinite-length ribbons.
Furthermore, we have introduced the following band-summed and band-mixing operators
\begin{eqnarray}
\hat{n}^B_{m\alpha\sigma}&=&\sum_{\tau=\pm} \hat{n}_{m\alpha\tau\sigma}=\sum_{\tau=\pm} \hat{\phi}^\dagger_{m\alpha\tau\sigma}\,\hat{\phi}_{m\alpha\tau\sigma}\\
\hat{n}^E_{m\sigma}&=&\sum_{\tau=\pm} \hat{n}_{m\tau\sigma}=\sum_{\tau=\pm} \hat{\psi}^\dagger_{m\tau\sigma}\,\hat{\psi}_{m\tau\sigma}\nonumber\\
\hat{P}_{m\sigma}&=&\sum_{\tau=\pm} \hat{P}_{m\tau\sigma}=\sum_{\tau=\pm} \hat{\psi}^\dagger_{m\tau\sigma}\,\hat{\psi}_{m\bar{\tau}\sigma}\nonumber
\end{eqnarray}
\noindent
where $\bar{\tau}=-\tau$. 

\subsection{Coulomb integrals}
\label{Subsection:Coulombintegrals}
\begin{figure}[ht]  \centering
  \includegraphics[width=\columnwidth]{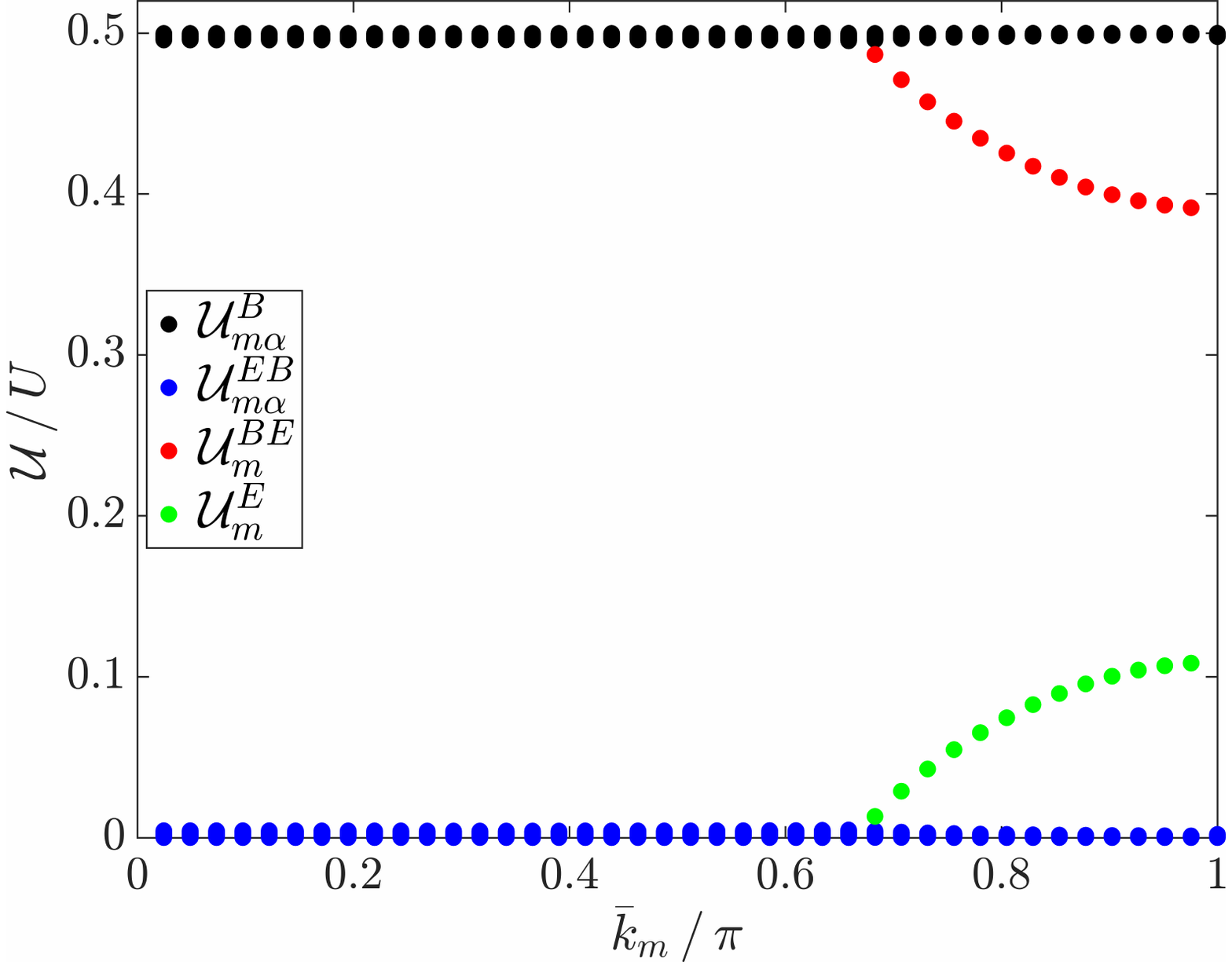}
  \caption{Coulomb integrals ${\cal U}^B_{m\alpha}$, ${\cal U}^{EB}_{m\alpha}$, ${\cal U}^{BE}_m$ and ${\cal U}^{E}_m$ in units of $U$ 
               (black, blue, red and green dots, respectively) as a function of the wave-number $\bar{k}_m$ for a rectangulene with dimensions 
               $(M_x,\,M_y) = (30,\,41)$.}
  \label{Figure:Uij}
\end{figure}

Using the approximate expressions in Eq. (\ref{Equation:sitenumber}), we decompose the interacting term in the Hamiltonian  into bulk, edge and crossed
contributions as follows:
\begin{eqnarray}
\hat{V}^{ee}&=&\hat{V}^\mathrm{B}+\hat{V}^\mathrm{E}+\hat{V}^\mathrm{BE}\\
\hat{V}^\mathrm{B}&=&\sum_{m m' \alpha \alpha'}\,U^B_{m\alpha,m'\alpha'}\,
\hat{n}^B_{m\alpha\uparrow}\hat{n}^B_{m'\alpha'\downarrow}\nonumber\\
\hat{V}^\mathrm{E}&=&\sum_{m m'}\,U^E_{m, m'}\,\left(\hat{n}^E_{m\uparrow}\hat{n}^E_{m'\downarrow}+
\hat{P}_{m\uparrow}\hat{P}_{m'\downarrow}\right)\nonumber\\
\hat{V}^\mathrm{BE}&=& \sum_{m m' \alpha \sigma}\,U^{BE}_{m\alpha,m'}\,
\hat{n}^B_{m\alpha\sigma}\hat{n}^E_{m'\bar{\sigma}}
\nonumber
\end{eqnarray}
\noindent
with $\bar{\sigma} = - \sigma$.
The Coulomb matrix elements are
\begin{eqnarray}
U^B_{m\alpha,m'\alpha'}&=&\frac{\cal C}{\Lambda^\phi_{m\alpha}\,\Lambda^\phi_{m'\alpha'}}\,\sum_{i=1,2}\,U_{m, m', i}^y\,U_{m\alpha,m'\alpha', i}^{B,x}\\
U^E_{m, m'}&=&\frac{\cal C}{\Lambda^\psi_m\,\Lambda^\psi_{m'}}\,\sum_{i=1,2}\,U_{m, m', i}^y\, U^{E,x}_{m, m', i}\nonumber\\
U^{BE}_{m\alpha,m'}&=&\frac{\cal C}{\Lambda^\phi_{m\alpha}\,\Lambda^\psi_{m'}}\,\sum_{i=1,2}\,U_{m, m', i}^y\, U^{BE,x}_{m \alpha,m',i}\nonumber
\end{eqnarray}
where
\begin{eqnarray}
{\cal C}&=&\frac{32\,U}{{\cal M}_x^2\,M_y^2}\\
U_{m\alpha,m'\alpha',i}^{B,x}&=&\sum_{R_x}\,\left(\phi_{m\alpha,i}^{\cal A}(R_x)\,\,\phi_{m'\alpha',i}^{\cal A}(R_x)\right)^2\nonumber\\
U^{E,x}_{m,m',i}&=&\sum_{R_x}\,\left(\psi_{m,i}^{\cal A}(R_x)\,\,\psi_{m',i}^{\cal A}(R_x)\right)^2\nonumber\\
U^{BE,x}_{m\alpha,m',i}&=&\sum_{R_x}\,\left(\phi_{m\alpha,i}^{\cal A}(R_x)\,\,\psi_{m',i}^{\cal A}(R_x)\right)^2\nonumber\\
U_{m, m',i}^y&=&\sum_{R_y}\,\left(f_{m,i}(R_y)\,\,f_{m',i}(R_y)\right)^2\nonumber
\end{eqnarray}
The above sums can be evaluated analytically. So after some straightforward but lengthy algebra we find that
\begin{eqnarray}
U^B_{m\alpha,m'\alpha'}&=&{\cal D}
\,\,\frac{\delta^+_{m m'}\,C^{B,+}_{m\alpha,m'\alpha'}+\delta^-_{m m'}\,C^{B,-}_{m\alpha,m'\alpha'}/
{\cal M}_x}{\Lambda^\phi_{m\alpha}\,\Lambda^\phi_{m'\alpha'}}\nonumber\\
U^E_{m, m'}&=&{\cal D}\,\,\frac{\delta^+_{m m'}\,C^{E,+}_{m,m'}+
\delta^-_{m m'}\,C_{m,m'}^{E,-}/{\cal M}_x}{G^1_m\,G^1_{m'}}\nonumber\\
U^{BE}_{m\alpha,m'}&=&2\,{\cal D}\,\,\,\frac{\delta^+_{m m'}\,C^{BE,+}_{m\alpha,m'}
+\delta^-_{m m'}\,C^{BE,-}_{m\alpha,m'}/{\cal M}_x}{\Lambda_{m\alpha}\,G^1_{m'}}\nonumber
\end{eqnarray}
Some intermediate steps of the above derivation can be found in the appendix. We have introduced the following short-hand notation to simplify the expressions above
\begin{eqnarray}
{\cal D}&=&\frac{U}{2\,{\cal M}_x\,M_y}\\
\delta_{m m'}^+&=&1+\frac{1}{2}\left(\delta_{m,m'}+\delta_{m,M_y} \delta_{m',M_y}\right)\nonumber\\
\delta_{m m'}^-&=&\frac{1}{2}\,\delta_{m+m',M_y}-\left(\delta_{m,M_y}+\delta_{m',M_y}\right)\nonumber
\end{eqnarray}
The formulae for the coefficients $C^B$, $C^E$ and $C^{BE}$ are rather cumbersome and we relegate them to 
the appendix. We have also found the important sum rules
\begin{eqnarray}
\frac{U}{2}&=&{\cal U}^B_{m\alpha}+{\cal U}^{EB}_{m\alpha}\\
                &=&{\cal U}^E_m+{\cal U}^{BE}_m\nonumber
\end{eqnarray}
where the summed Coulomb integrals      
\begin{eqnarray}          
{\cal U}^B_{m\alpha}&=&\sum_{m'\alpha'}\,U^B_{m\alpha, m'\alpha'}\\
{\cal U}^{BE}_{m\alpha}&=&\sum_{m'}U^{BE}_{m\alpha, m'}\nonumber
\label{Equation:sumrule}
\end{eqnarray}
give a measure of the relevance of bulk and edge contributions to the renormalization of the dispersion relation of bulk states. 
Conversely, the sums
\begin{eqnarray}
{\cal U}^E_m&=&\sum_{m'}\,U^E_{m, m'}\\
{\cal U}^{EB}_m&=&\sum_{m' \alpha'}U^{BE}_{m'\alpha', m}\nonumber
\end{eqnarray}
give a measure of the contributions of edge and bulk states to the renormalization of the dispersion relation of edge states.
We plot the values of these different Coulomb integrals as a function of $\bar{k}_m$ in Figure \ref{Figure:Uij}, 
for a rectangulene with dimensions $(M_x,\,M_y)=(30,\,41)$, e.g.: 12.8 $\times$ 10.1 nm. Interestingly,
${\cal U}^B_{m\alpha}$ is almost constant and approximately equal to $0.5\,U$ for all $\bar{k}_y$
wave-numbers. Similarly, ${\cal U}^{EB}_m$ is rather small but non-zero. In contrast, ${\cal U}^E_m$ and
${\cal U}^{BE}_{m\alpha}$ are both different from zero and feature a strong dependence with $\bar{k}_m$.
The important message here is the unexpected large contribution of bulk states 
to the edge-state dispersion relation.

\subsection{Bulk and edge Hubbard Hamiltonians}
We rewrite the Hubbard Hamiltonian is its final form
\begin{eqnarray}
\hat{H}=\hat{H}^B+\hat{H}^E+\hat{V}^\mathrm{BE}
\end{eqnarray}
where the bulk and edge Hamiltonians are as follows
\begin{eqnarray}
\hat{H}^B&=&\sum_{m\alpha\tau\sigma} \tau\,\epsilon^B_{m\alpha}\,\hat{n}^B_{m\alpha\tau\sigma}+
\sum_{m \alpha,m'\alpha'}\,U^B_{m\alpha,m'\alpha'}\,\hat{n}^B_{m\alpha\uparrow}\hat{n}^B_{m'\alpha'\downarrow}\nonumber\\\\
\hat{H}^E&=&\sum_{m\tau\sigma} \tau\,\epsilon^E_m\,\hat{n}^E_{m\tau\sigma}+
\sum_{m m'}\,U^E_{m m'}\,\left(\hat{n}^E_{m\uparrow}\hat{n}^E_{m'\downarrow}+
\hat{P}_{m\uparrow}\hat{P}_{m'\downarrow}\right)\nonumber
\end{eqnarray}

\subsection{Occupations and magnetization}
The sublattice- and spin-resolved total occupations are 
\begin{eqnarray}
\left(\begin{matrix}{\cal N}_\sigma^{\cal A}\\{\cal N}_\sigma^{\cal B}\end{matrix}\right)
=\frac{1}{2}\sum_{m\alpha}\left(\begin{matrix}\braket{\hat{n}^B_{m\alpha\sigma}}\\\braket{\hat{n}^B_{m\alpha\sigma}}\end{matrix}\right)+
\frac{1}{2}\sum_m\left(\begin{matrix}\braket{\hat{n}^E_{m\sigma}}+\braket{\hat{P}_{m\sigma}}\\\braket{\hat{n}^E_{m\sigma}}-\braket{\hat{P}_{m\sigma}}\end{matrix}\right)\nonumber\\
\end{eqnarray}
where ${\cal N}_\sigma={\cal N}_\sigma^{\cal A}+{\cal N}_\sigma^{\cal B}$ and ${\cal N}={\cal N}^B+{\cal N}^E={\cal N}_\uparrow+{\cal N}_\downarrow$.
The magnetization per edge state can have contributions from both bulk and edge states
\begin{eqnarray}
m&=&\frac{{\cal N}_\uparrow-{\cal N}_\downarrow}{N^\mathrm{edge}}=m^B+m^E\\
m^B&=&\frac{1}{N^\mathrm{edge}}\,\sum_{m\alpha\sigma}\,\sigma\braket{\hat{n}^B_{m\alpha\sigma}}\nonumber\\
m^E&=&\frac{1}{N^\mathrm{edge}}\,\sum_{m\sigma}\,\sigma\braket{\hat{n}^E_{m\sigma}}\nonumber
\end{eqnarray}
although for low enough doping levels, only edge states contribute. 
The sublattice-unbalanced magnetization
\begin{eqnarray}
m^\mathrm{st}=\frac{1}{N^\mathrm{edge}}\,\sum_\sigma \,\sigma\,\left({\cal N}^{\cal B}_\sigma-{\cal N}^{\cal A}_\sigma\right)=
\frac{1}{N^\mathrm{edge}}\,\sum_{m\sigma}\,\sigma\,\braket{\hat{P}_{m\sigma}}\nonumber\\
\end{eqnarray}
is a measure of the staggered magnetization across edges. Fernandez-Rossier introduced \cite{Rossier2008}
a {\it spin dipole operator} which is in essence the band-mixing operator $\hat{P}_{m\sigma}$ defined in the present article.

\section{Mean-field solutions of the Hubbard model of a graphene rectangulene}
\label{Section:meanfieldsolutions}
Most Mean-Field treatments of this model perform a collinear spin decomposition of the four-fermion Coulomb operator 
directly in the site-basis as follows
\begin{eqnarray}
\hat{n}_{{\bf R} i \uparrow} \hat{n}_{{\bf R} i \downarrow}\approx n_{{\bf R} i \uparrow} \hat{n}_{{\bf R} i \downarrow} + \hat{n}_{{\bf R} i \uparrow} n_{{\bf R} i \downarrow}-n_{{\bf R} i \uparrow} n_{{\bf R} i \downarrow}
\end{eqnarray}
with the spin-dependent site occupations being defined as $n_{{\bf R} i \sigma} =\braket{\hat{n}_{{\bf R} i \sigma}}$.
This approach results in a one-body Hamiltonian that depends on the occupations $n_{{\bf R} i \sigma}$, 
that must be determined via a set of coupled self-consistency equations. These equations have been solved numerically many
times in the past for different graphene nano-structures, but never analytically. We cite here references 
\cite{Rossier2008} and \cite{Jung2009a,Jung2009b} among many other.
The numerical difficulty lies on the fact that the number of different occupations is twice the number of atoms in the 
device.
For infinite-length ribbons, traslational symmetry along the ribbon axis means that a unit cell can be defined so that 
the number of different occupations is just twice the number of atoms in the unit cell. 
However, no traslational symmetry exists for graphene rectangulenes, or more generally for graphene flakes so that 
number of sites occupations grows like the square of the lateral size. All in all, it is not possible to simulate 
experimental-sized rectangulenes or flakes  with today's numerical capabilities. 
\subsection{Bulk states}
We apply first a mean-field approximation to the bulk piece of the Hamiltonian $\hat{H}^B$, where 
$n^B_{m\alpha\tau\sigma}=\braket{\hat{n}^B_{m\alpha\tau\sigma}}$. We find
\begin{eqnarray}
\hat{H}^B_{MF}&=&\sum_{m\alpha\tau\sigma}\,\xi^B_{m\alpha\tau\sigma}\,\hat{n}^B_{m\alpha\tau\sigma}
\\
\xi^B_{m\alpha\tau\sigma}&=&\tau\,\epsilon^B_{m\alpha}+\sum_{m'\alpha'}\,U^B_{m\alpha, m'\alpha'}\,
n^B_{m'\alpha'\bar{\sigma}}+\sum_{m'}U^{BE}_{m\alpha, m'}\,n^E_{m'\bar{\sigma}}\nonumber
\end{eqnarray}
The bulk contribution to the total energy is
\begin{eqnarray}
E_T^B&=&\sum_{m\alpha\tau\sigma}\,\xi^B_{m\alpha\tau\sigma}\,n_{m\alpha\tau\sigma}-E_\mathrm{dc}^B\\
E_\mathrm{dc}^B&=&\sum_{m m' \alpha\alpha'}\,U^B_{m\alpha, m'\alpha'}\,n^B_{m\alpha\uparrow}\,n^B_{m'\alpha'\downarrow}\nonumber
\end{eqnarray}
We define the {\it edge-only doping} regime as the regime where the doping is low enough that the added (extracted) 
electrons are allocated into (picked from ) the edge states. As 
a consequence, bulk states will always be fully filled for the valence band, $n_{m\alpha -\sigma}=1$, and 
completely empty, $n_{m\alpha +\sigma}=0$, for the conduction band.
Hence the number of electrons residing in bulk states is ${\cal N}^B=2(\,M_x\,N-N^\mathrm{edge})$, where the number of
edge states $N^\mathrm{edge}$ is given in Eq. (\ref{Equation:nedge}). The bulk dispersion relation simplifies to
\begin{eqnarray}
\xi^B_{m\alpha\tau\sigma}&=&\tau\,\epsilon^B_{m\alpha}+{\cal U}^B_{m\alpha}+\sum_{m'}\,U^{BE}_{m\alpha, m'}\,n^E_{m'\bar{\sigma}}
\label{Equation:simplification}
\end{eqnarray}
And the contribution of bulk states to the rectangulene total energy is
\begin{eqnarray}
E^B_T=\sum_{m\alpha}\left(\xi^B_{m\alpha - \uparrow}+\xi^B_{m\alpha - \downarrow}-{\cal U}^B_{m\alpha}\right)
\end{eqnarray}
We also have that the bulk contribution to the magnetization is identically zero, $m^B=0$.

The apparently anodyne result in Eq. (\ref{Equation:simplification}) when paired with the sum rule in Eq. 
(\ref{Equation:sumrule}) amounts to a huge simplification that enables us to carry out calculations for huge rectangulenes,  
because the number of Coulomb matrix elements to be calculated and stored is reduced from $M_x^2\,N^2$ to $M_x\,N N^{edge}$.  

\subsection{Edge states}
We perform now a mean-field approximation to the edge piece of the Hamiltonian so that 
$n^E_{m\tau\sigma}=\braket{\hat{n}^E_{m\tau\sigma}}$ and $P_{m\tau\sigma}=\braket{\hat{P}_{m\tau\sigma}}$. 
We also have the band-summed relationships 
$n^E_{m\sigma}=n^E_{m - \sigma}+n^E_{m + \sigma}$ and 
$P_{m\sigma}=P_{m - \sigma}+P_{m + \sigma}$. The edge mean-field Hamiltonian becomes
\begin{eqnarray}
\hat{H}^E_{MF}&=&\sum_{m\tau\sigma}\,\xi^E_{m\tau\sigma}\,\hat{n}^E_{m\tau\sigma}+
\sum_{m\sigma}\,U^E_{m,m'}\,P_{m'\bar{\sigma}}\,\hat{P}_{m\sigma}\\
\xi^E_{m\tau\sigma}&=&\tau\,\epsilon^E_{m}+{\cal U}^{BE}_m+\sum_{m'}\,U^E_{m,m'}\,n^E_{m'\bar{\sigma}}\nonumber
\end{eqnarray}
Diagonalization of the mean-field edge Hamiltonian provides us with the edge eigen-energies $\xi^E_{m\tau\sigma}$ and 
edge eigen-states. The edge states' occupations are determined by the equation of state
\begin{eqnarray}
n^E_{m\sigma}&=&n_F(\xi^E_{m + \sigma})+n_F(\xi^E_{m - \sigma})
\end{eqnarray}
where $n_F$ is  the Fermi function. The order parameter $P_{m\sigma}$ is determined by solving the self-consistency
equations adequate to each mean-field solution, as discussed below.

We introduce the edge doping $\delta^E$ by taking half-filling as a reference:
\begin{eqnarray}
\delta^E={\cal N}^E-2\,N^\mathrm{edge}= \sum_{m\sigma} n^E_{m\sigma}-2\,N^\mathrm{edge}
\end{eqnarray}
The edge contribution to the rectangulene total energy is determined by the equation
\begin{eqnarray}
E^E_T&=&\sum_{m\tau\sigma}  \xi^E_{m \tau \sigma}\, n^E_{m \tau\sigma} -E_\mathrm{dc}^E\\
E_\mathrm{dc}^E&=&\sum_{m m'}\,U^E_{m,m'}\,\left(n^E_{m\uparrow}\,n^E_{m'\downarrow}+P_{m\uparrow}\,P_{m'\downarrow}\right)+
\sum_m\,{\cal U}^{BE}_m\,n^E_m\nonumber
\end{eqnarray}

\subsection{Paramagnetic solution}
\label{Subsection:pmsolution}
The PM mean-field solution is found by setting $n^E_{m\tau\uparrow}=n^E_{m\tau\downarrow}=n^E_{m\tau}/2$ and 
$P_{m\tau\sigma}=0$. Then the edge and bulk dispersion relations do not depend on the spin degree of freedom
\begin{eqnarray}
\xi^E_{m\tau}&=&\tau\,\epsilon^E_{m}+{\cal U}^{BE}_m+\frac{1}{2}\sum_{m'}U^E_{m m'}\,n^E_{m'}\\
\xi^B_{m\alpha\tau}&=&\tau\,\epsilon^B_{m\alpha}+{\cal U}^B_{m\alpha}+\frac{1}{2}\sum_{m'}U^{BE}_{m\alpha m'} n^E_{m'}\nonumber
\end{eqnarray}
Finally, the edge double-counting contribution to the total energy is 
\begin{eqnarray}
E_\mathrm{dc}^E&=&\frac{1}{4}\,\sum_{m m'}\,U^E_{m,m'}\,n^E_m\,n^E_{m'}+\sum_m\,{\cal U}^{BE}_m\,n^E_m
\end{eqnarray}

\subsection{Ferromagnetic solution}
\label{Subsection:fmsolution}
The FM mean-field solution is found by setting $n^E_{m\tau\uparrow}\neq n^E_{m\tau\downarrow}$ and 
$P_{m\tau\sigma}=0$. Then the edge and bulk dispersion relations are
\begin{eqnarray}
\xi^E_{m\tau\sigma}&=&\tau\,\epsilon^E_{m}+{\cal U}^{BE}_m+\sum_{m'}U^E_{m m'}\,n^E_{m'\bar{\sigma}}\\
\xi^B_{m\alpha\tau\sigma}&=&\tau\,\epsilon^B_{m\alpha}+{\cal U}^B_{m\alpha}+\sum_{m'}U^{BE}_{m\alpha m'} 
n^E_{m'\bar{\sigma}}\nonumber
\end{eqnarray} 
and the edge double-counting term is
\begin{eqnarray}
E_\mathrm{dc}^E&=&\sum_{m m'}\,U^E_{m,m'}\,n^E_{m\uparrow}\,n^E_{m'\downarrow}+\sum_m\,{\cal U}^{BE}_m\,n^E_m
\end{eqnarray}

As shown in our previous work \cite{Amador2023}, a FM coupling between edge states only occurs for long enough rectangulenes,
that is when the ${\cal U}_m^E$ term dominates over the $\epsilon_m^E$ term. If this is the case,
the equation of state can be solved at zero temperature and arbitrary doping $\delta^E$. 
For positive doping values, the majority spin states $\uparrow$ are completely filled with $n^E_{m\uparrow}=2$. As ${\cal U}_m^E$ 
increases with $m$ (see Fig. \ref{Figure:Uij}), the minority spinstates $\downarrow$ are filled following the order of $m$.
As $m$ runs from $m^E_{\mathrm{min}}=M_y-N^\mathrm{edge}$ to $m^E_{\mathrm{max}}=M_y-1$, we define a critical $m^E_c = m^E_{\mathrm{min}}+\mathrm{floor}(\delta^E/2)$,  so:
\begin{eqnarray}
n^E_{m^E_{\mathrm{min}}:m^E_c-1,\downarrow} &=&2\\
n^E_{m^E_c,\downarrow}& =&\delta^E-2\,\mathrm{floor}(\delta^E/2)\nonumber\\
n^E_{m=m^E_c+1:m^E_{\mathrm{max}},\downarrow}& =&0\nonumber
\end{eqnarray}
The dispersion relations can then be written explicitly as
\begin{eqnarray}
\xi^E_{m\tau\downarrow}&=&\tau\,\epsilon_m+\frac{U}{2}+{\cal U}^E_m\\
\xi^E_{m\tau\uparrow}&=&\tau\,\epsilon_m+\frac{U}{2}+\left(\sum_{m'=m^E_{\mathrm{min}}}^{m^E_c-1}-\sum_{m'=m_c+1}^{m^E_\mathrm{max}}\right)
\,U^E_{m,m'} + \nonumber\\&&+ U^E_{m,m^E_c}\,(n^E_{m^E_c,\downarrow} - 1)\nonumber\\
\xi^B_{m\alpha\tau\downarrow}&=&\tau\,\epsilon^B_{m\alpha}+\frac{U}{2}+{\cal U}^{BE}_{m \alpha} \nonumber\\
\xi^B_{m\alpha\tau\uparrow}&=&\tau\,\epsilon^B_{m\alpha}+\frac{U}{2}+
\left(\sum_{m'=1}^{m_c}-\sum_{m'=m_c+2}^{N^\mathrm{edge}}\right)\,U^{BE}_{m\alpha,m'}+\nonumber\\
&&+U^{BE}_{m\alpha,m^E_c}\,(n^E_{m^E_c,\downarrow} - 1)\nonumber
\end{eqnarray}
Similarly, the FM total energy has the following analytical expression
\begin{eqnarray}
E_T&=&\sum_{m\alpha\sigma}\,\xi^B_{m\alpha - \sigma}+\sum_{m\tau\sigma}\,\xi^E_{m\tau\sigma}\,n^E_{m\tau\sigma}-
\frac{U}{2}(1+\delta^E)\nonumber\\&&
-\sum_{m\alpha}{\cal U}^{BE}_{m\alpha}-\sum_{m m'}U^E_{m,m'}\,n^E_{m'\downarrow}
\end{eqnarray}

\subsection{Antiferromagnetic solution}
\label{Subsection:afmsolution}
The AFM solution is selected by letting the order parameter $P_{m\tau\sigma}$ be different from zero.
In this case the edge Hamiltonian looks 
\begin{eqnarray}
\hat{H}^E_{MF}&=&\sum_{m\tau\sigma}\,(\tau\,\epsilon_m^E+H_{m\sigma})\,
\hat{n}^E_{m\tau\sigma}+\sum_{m\sigma}\Delta_{m\sigma}\hat{P}_{m\sigma}\\
H_{m\sigma}&=&{\cal U}^{BE}_m+\sum_{m'}\,U^E_{m,m'}\,n^E_{m'\bar{\sigma}}\nonumber\\
\Delta_{m\sigma}&=&\sum_{m'}U^E_{m,m'}\,P_{m'\bar{\sigma}}\nonumber
\end{eqnarray}
The above Hamiltonian can be rewritten in the following BCS form
\begin{eqnarray}
\sum_{m\sigma}\,\left(\begin{matrix}\hat{\psi}_{m+\sigma}^\dagger \,\hat{\psi}_{m -\sigma}^\dagger\end{matrix}\right)\,
\left(\begin{matrix}H_{m\sigma}+\epsilon_m&\Delta_{m\sigma}\\
\Delta_{m\sigma}&H_{m\sigma}-\epsilon_m\end{matrix}\right)\,
\left(\begin{matrix}\hat{\psi}_{m+\sigma}\\\hat{\psi}_{m -\sigma}\end{matrix}\right)\nonumber\\
\end{eqnarray}
and can be diagonalized by a Bogoliubov transformation so that
\begin{eqnarray}
\hat{H}^E_{MF}&=&\sum_{m\alpha\sigma}\,\xi^E_{m\tau\sigma}\,\hat{\gamma}^\dagger_{m\alpha\sigma}\,\hat{\gamma}_{m\alpha\sigma}\\
\xi^E_{m\tau\sigma}&=&\xi^E_{m\tau}=H_{m\sigma}+\tau\,R_m=H_{m\sigma}+\tau\,\sqrt{\epsilon_m^2+\Delta_m^2}\nonumber
\end{eqnarray}
where we have considered $P_{m\sigma}=\sigma P_m$, so $\Delta_{m\sigma}^2$=$\Delta_{m}^2$ independent of the spin.
Therefore, the edge eigen-energies are spin-degenerate. The eigen-state operators and Bogoliubov coherence factors are
\begin{eqnarray}
\left(\begin{matrix}\hat{\gamma}_{m + \sigma}\\\hat{\gamma}_{m - \sigma}\end{matrix}\right)&=&
\left(\begin{matrix}u_{m\sigma}^+ &\mathrm{sign}(\Delta_{m\sigma})\,u_{m\sigma}^-
\\-\mathrm{sign}(\Delta_{m\sigma})\,u_{m\sigma}^- &u_{m\sigma}^+\end{matrix}\right)
\left(\begin{matrix}\hat{\psi}_{m + \sigma}\\\hat{\psi}_{m - \sigma}\end{matrix}\right)\nonumber\\
u_{m\sigma}^{\pm}&=&\frac{1}{\sqrt{2}}\,\left(1\pm\frac{\epsilon^E_m}{R_m}\right)^{1/2}
\end{eqnarray}
Finally, the order parameter can be determined by the conventional BCS  self-consistency equations 
\begin{eqnarray}
P_{m\sigma}&=&\frac{\Delta_{m\sigma}}{R_m}\left(n_F(\xi^E_{m + \sigma})-n_F(\xi^E_{m - \sigma})\right)
\label{Equation:selfconsistency}
\end{eqnarray}

\begin{figure}[ht]  \centering
  \includegraphics[width=\columnwidth]{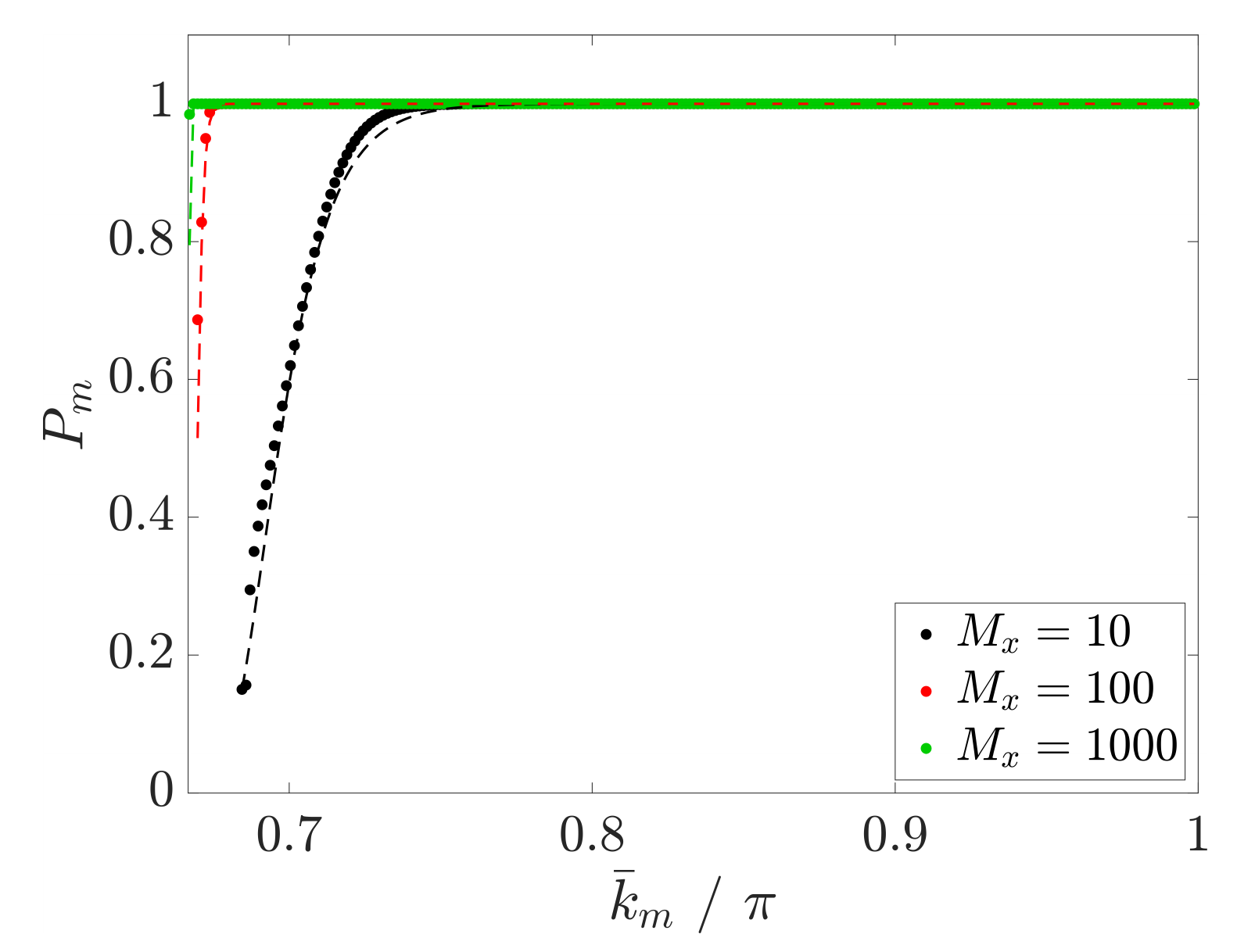}
  \caption{Zero-temperature AFM order parameter $P_m$ of an undoped rectangulene of width $N=1501$, 
               that corresponds to 184,7 nm and hosts  500 states.  Black, red and green dots correspond to lengths 
               $M_x=10,\,100$ and $1000$ (4.26, 42,6 and 426 nm, respectively). Dashed lines show the fitting of the results to $\tanh^4{\left ( {\cal M}_x \left ( \frac{\bar{k}_m}{\pi}-\frac{2}{3} \right ) \right ) }$} 
  \label{Figure:orderparameter}
\end{figure}

Just like for the FM solution, the AFM coupling between edge states only occurs for long enough rectangulenes \cite{Amador2023}.
If this is the case, the equation of state can be solved at zero temperature for the AFM solution similarly to the 
FM case. We find that the occupations are $n^E_{m, - \sigma}=1$ and
\begin{eqnarray}
n^E_{m^E_{\mathrm{min}}:m^E_c-1,+ \sigma} &=&1 \\
n^E_{m^E_c,+ \sigma}& =&\frac{\delta^E}{2}-\mathrm{floor}(\delta^E/2)\nonumber\\
n^E_{m^E_c+1:m^E_{\mathrm{max}},+ \sigma}& =&0\nonumber
\end{eqnarray}  
As a consequence, 
\begin{eqnarray}
H_{m\sigma}=\frac{U}{2}+\sum_{m'=m^E_{\mathrm{min}}}^{m_c-1}\,U^E_{m,m'}+U^E_{m,m_c}\, n^E_{m_c + \bar{\sigma}}
\end{eqnarray} 

The order parameter $P_{m}(M_x,\,M_y)$ must be determined by solving numerically equation 
(\ref{Equation:selfconsistency}). We find that the absolute value of the order parameter does not depend 
on the rectangulene's width $M_y$.
We plot $P_m(M_x)$ as a function of $\bar{k}_m$ in Figure \ref{Figure:orderparameter} for several undoped  
rectangulene's lengths  $M_x$, and at zero temperature. We have tested several functional forms to fit these curves, and have found 
that $P_m$ can be reasonably approximated by the function $\tanh^4{\left ( {\cal M}_x \left ( \frac{\bar{k}_m}{\pi}-\frac{2}{3} \right ) \right ) }$ 
to a high accuracy.  Furthermore, for large enough $M_x > 50-100$, the order parameter of a doped
rectangulene can be approximated as follows:
\begin{eqnarray}
P_{m^E_{\mathrm{min}}:m^E_c-1} &=& 0\\
P_{m^E_c}&=&1+\mathrm{floor}(\delta^E/2)-\delta^E/2\nonumber\\
P_{m^E_c+1:m^E_{\mathrm{max}}}&=& 1\nonumber
\end{eqnarray}
so that
\begin{eqnarray}
\Delta_{m\sigma}=-\sigma\,\left(P_{m_c^E}\,U^E_{m, m_c^E}+\sum_{m'=m_c^E+1}^{m^E_{\mathrm{max}}}\,U^E_{m,m'}\right)
\end{eqnarray}

Fernandez-Rossier proposed a phenomenological BCS-like description of band mixing \cite{Rossier2008} that is 
consistent with our results above. Similarly, MacDonald and coworkers developed a phenomenological BCS model of 
inter-edge mixing that is also consistent with our results \cite{Jung2009b}.

\begin{figure}[ht]  \centering
  \includegraphics[width=\columnwidth]{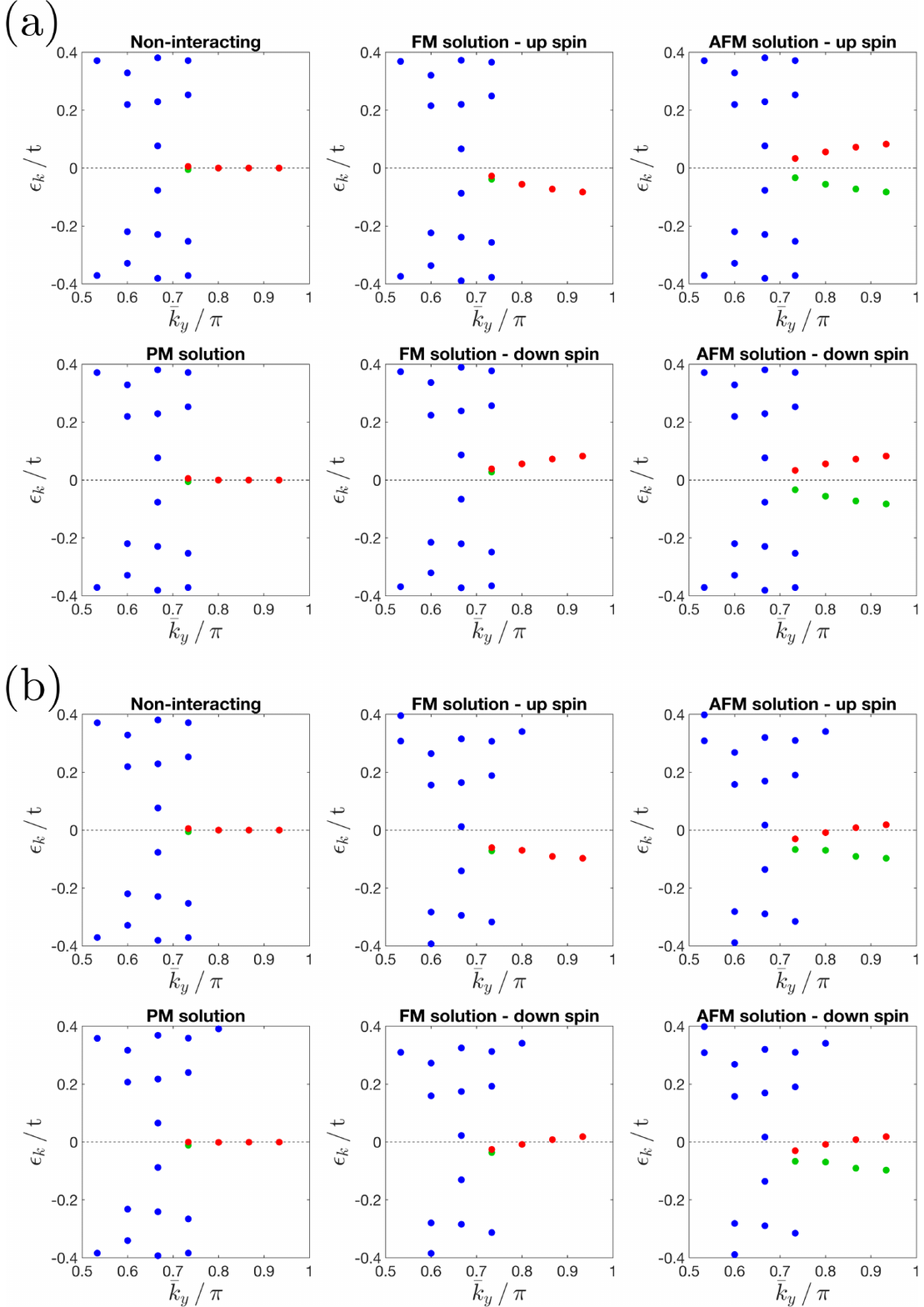}
  \caption{Electronic structure of a $(\,M_x,\,M_y\,) = (10,15)$, $N=9$ rectangulene as a function of the $k_y$ 
           wave-number
           at an edge filling (a) $\delta^E=0$ and (b) $\delta^E=4$ electrons. This rectangulene 
           has dimensions 4.3 nm $\times$ 3.7 nm, and hosts 8 edge states.  The left column plots 
           the non-interacting (top) and mean-field PM electronic structure. The central/right columns 
           plot the mean-field FM/AFM electronic structure for spin-up (top) and spin-down (bottom).
           Blue dots correspond to bulk states; red/green dots correspond to $\tau=+1\,/\,-1$ edge states.}
  \label{Figure:bandsMx10My15}
\end{figure}

\begin{figure}[ht]  \centering
  \includegraphics[width=\columnwidth]{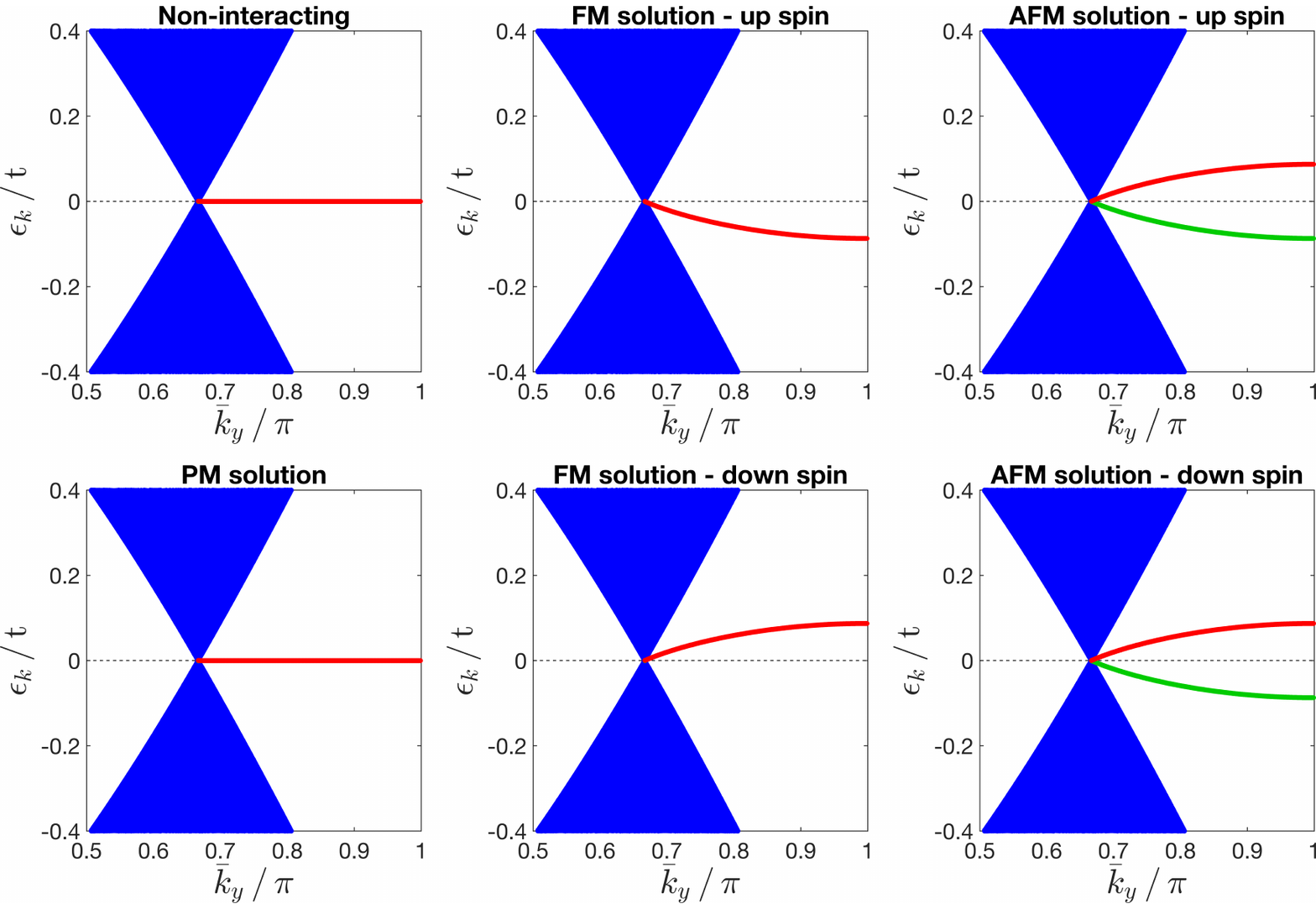}
  \caption{Same as in Figure \ref{Figure:bandsMx10My15}, now for a $(\,M_x,\,M_y\,) = (950,901)$ , $N=1801$
           rectangulene whose dimensions are 404.7 nm $\times$ 221.6 nm, and host 600 edge states. This 
           rectangulene is made of about 3.5 million atoms.}
  \label{Figure:bandsMx1000My1001}
\end{figure}

\subsection{Beyond the  {\it edge-only doping} regime}
\label{Subsection:beyond}
The {\it edge-only doping} regime is too restrictive an approximation in several instances. Examples are 
gate- or voltage-biased finite-length 7-AGNR \cite{Zhang2023b} or bulk-size rectangulenes where the Dirac-point
gap is negligible. Fortunately, the approximation can be released to include low-lying bulk states within the
self-consitency procedure. This can be achieved by choosing a small energy cutoff $E_c$ so that occupations of 
bulk states with energies $|\xi^B_{m\alpha\tau\sigma}-\mu|$ smaller than $E_c$ are determined self-consistently, while 
those above this energy cutoff are frozen at 0 or 1.

\subsection{Dispersion relations}
The analysis above opens the door to determine easily the electronic structure on rectangulenes of any size in
the {edge-only doping} regime. We will take below a hopping integral $t=2.7$ eV, and a Hubbard-$U$ parameter equal 
to $0.8\,t$.

A first simple example of a short $N=29$ rectangulene is shown in Figure \ref{Figure:bandsMx10My15}. This rectangulene 
hosts 8 edge states, that can be easily spotted in the figure. We find the AFM phase to be more stable at zero doping
than the FM phase by 2.42 meV, while they are essentially degenerate when doped with 4 electrons. We find that 
four-electron doping leaves the rectangulene still in the {\it edge-only} doping regime. This electronic structure
follows the well-known trends of undoped and doped infinite-length zigzag GNRs \cite{Jung2009b}. However, the
states here might be better regarded as molecular orbitals rather than Bloch states, and the figure shows explicitly 
the discrete spectrum of mean-field eigen-energies. 

A second example is shown in Figure \ref{Figure:bandsMx1000My1001}, that corresponds to a bulk-like undoped graphene sheet 
{\it with edges}. This example has been chosen to illustrate the difference between periodic versus open boundary conditions 
for graphene, because graphene edges host edge states that should be discarded lightly. 
The dispersion relation in this case features not only the bulk Dirac cone, but also the quasi-continuum 1-dimensional 
spectrum corresponding to the edge branches. These edge branches cannot obtained using periodic boundary conditions;
The Dirac spectrum cannot be gotten by simulating graphene ribbons. We find here that the 
energy gap at the bulk Dirac point is smaller than the lowest edge eigen-energy, so that the {edge-only doping} regime
does not exist for this rectangulene.

Overall, we find that the edge states are always double-degenerate. There exist branch-degeneracy ($\tau=\pm$) 
but spin-degeneracy lifting for the FM solution. In contrast, there is spin-degeneracy but branch-degeneracy
lifting for the AFM solution.

\subsection{Addition energies}
\label{Subsection:additionenergies}
Recently, single $M_x=5$, $N=9$ rectangulenes have been deposited onto ultraclean 
graphene nanogaps, and the differential conductance as a function of both bias and gate voltage has been measured.
Neat sequences of Coulomb blockade diamonds have been observed \cite{Niu2023,Zhang2023a,Zhang2023b}, whereby
the device addition energies have been extracted.

The addition energy of a rectangulene having a total of ${\cal N}$ electrons is
\begin{eqnarray}
E^{add}({\cal N})=E_T({\cal N}+1)+E_T({\cal N}-1)-2\,E_T({\cal N})
\end{eqnarray}
Our exact solution enables us to compute easily these addition energies for arbitrary dopings at the mean field 
level. The 
{\it edge-only doping} approximation restricts the validity of the calculations to low dopings and 
lengths $M_x$ sufficiently small that bulk states have all higher energies than the edge states to be addressed.
This approximation can however be released easily as explained in section \ref{Subsection:beyond} 
above.

We have checked that $E^{add}$ depends on the rectangulene's width $M_y$ but not on its length $M_x$, 
because of the finite extent of edge states, as discussed in our previous article. 
The FM/AFM solutions have branch/spin degeneracy meaning that
$E^{add}({\cal N}=\mathrm{even})=0$, and we have checked that this is the case. We then find that
\begin{eqnarray}
E^{add}(\delta^E&=&\mathrm{odd})=\xi^E_{m+1\tau}-\xi^E_{m\tau}
\end{eqnarray}
for the AFM solution, while $\tau$ is replaced by $\sigma$ for the FM solution,
in agreement with Koopman's theorem \cite{Koopman34,Janak78}.
Additionally, we find that $E^{add}$ is the same for the FM and the AFM solutions.
We list in Table \ref{Table:table1} the addition energies of a rectangulene with $M_y=13$, $N=13$ 
(e.g.: width 3.2 nm), 
that can host up to 8 electrons in edge states. The {\it edge-only doping} regime restricts in this case the 
rectangulene's lengths to values $M_x<40$, corresponding to lengths of about 17 nm. 

\begin{table}[ht]
\small
  \caption{Addition energies of a rectangulene of width $M_y=13$.}
  \begin{tabular*}{0.48\textwidth}{@{\extracolsep{\fill}}ccccccccc}
     \hline
     $\delta^E$                & 0  & 1 & 2  & 3 &  4 & 5 &  6 & 7 \\
     \hline
     $E^{add}$ (meV) & 82 & 0 & 86 & 0 & 59 & 0 & 37 & 0 \\
    \hline
  \end{tabular*}
\label{Table:table1}
\end{table}

\begin{figure}[ht]  \centering
  \includegraphics[width=\columnwidth]{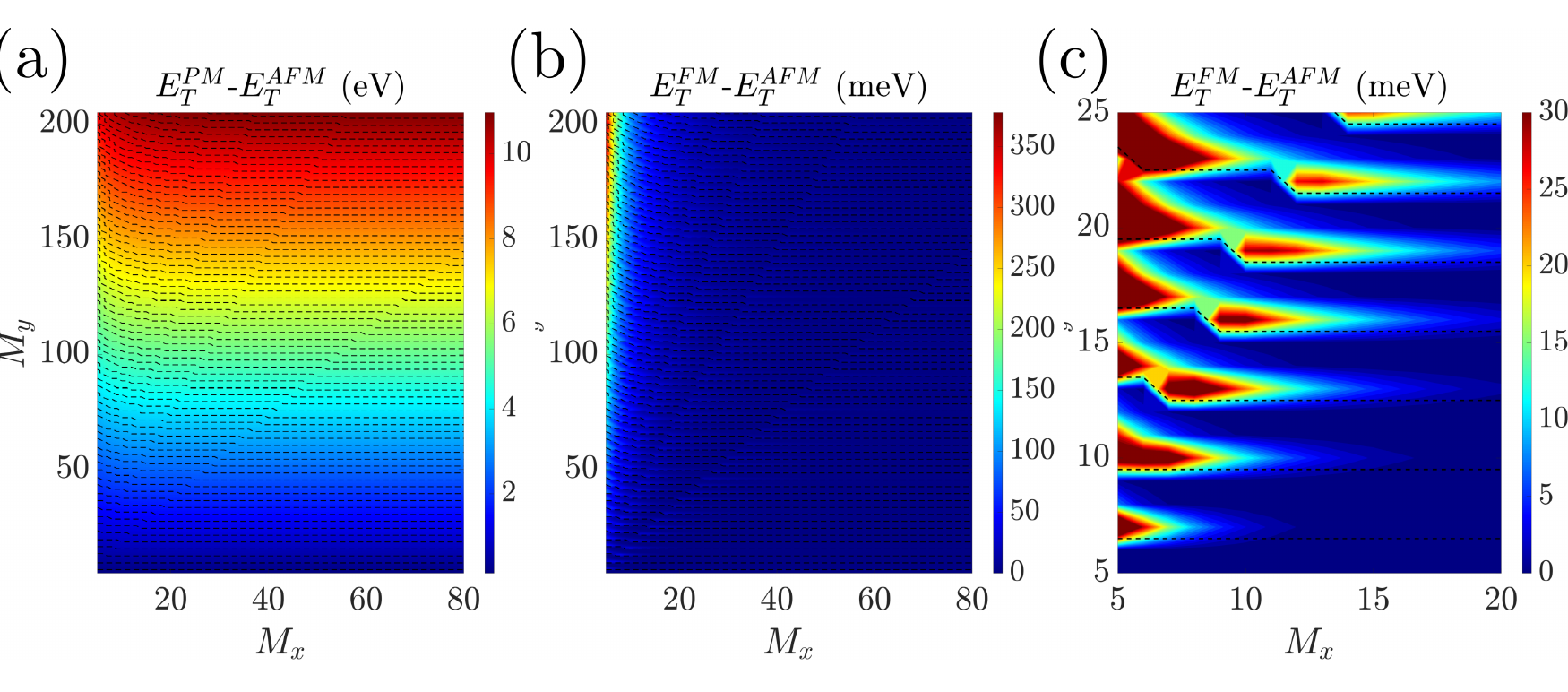}
  \caption{Energy differences (a) $E_T^{PM}-E_T^{AFM}$ and (b) $E_T^{FM}-E_T^{AFM}$ for undoped rectangulenes in a mesh of $M_x$ and $M_y$ values. 
  Dashed lines indicate the appearance of new edge states. Panel (c) shows $E_T^{FM}-E_T^{AFM}$ in a smaller mesh with more detail. }
  \label{Figure:ediffs}
\end{figure}

\subsection{Energy differences among phases}
Jung and MacDonald analyzed phase stabilities of narrow infinite-length zigzag GNRs as a function of 
doping\cite{Jung2009b}. We discuss here energy differences of undoped rectangulenes as a function of width and 
length (see Fig. \ref{Figure:ediffs}). Overall, we find that the magnetic energy, measured as the energy difference $E_T^{PM}-E_T^{AFM}$ is
positive as it should \cite{Hubbard63,Penn66} and roughly independent of $M_x$, especially for $M_x$ larger than about 20, as shown in Fig. 7 (a). This is 
expected because for large enough $M_x$, the tails of the edge wavefunctions decay enough that tails at opposite edges do not overlap. 
The magnetic energy is in contrast roughly proportional to $M_y$, e.g.: to the number of edge states. We also find that the
energy difference among the FM and AFM phases $E_T^{FM}-E_T^{AFM}$ is positive but decays quickly with $M_x$ so that the 
edge states at opposite edges become independent for $M_x$ larger than about 40-60, as shown in Fig. 7 (b). 
For the $M_y$ dependence, we find that $E_T^{FM}-E_T^{AFM}$ presents oscillations related to the
change in the number of edge states of the system, as shown in Fig. 7 (c).

\begin{figure}[ht]  \centering
  \includegraphics[width=\columnwidth]{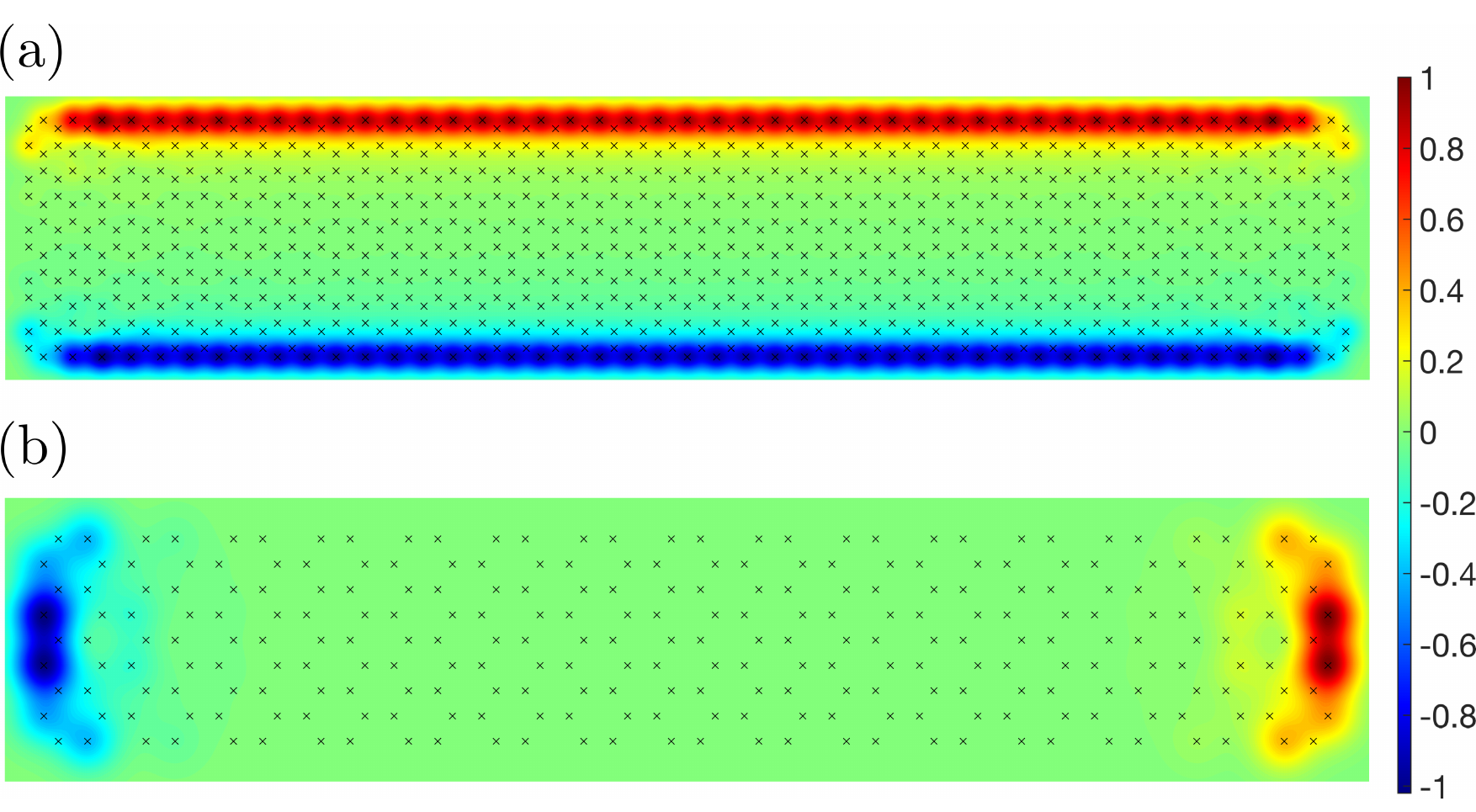}
  \caption{Edge magnetization for a rectangulene with (a) $(M_x, \,M_y)=(5,\,46)$, $N=91$ 
           whose size is 2.1 nm $\times$ 11.3 nm, and hosts 26 edge states; (b)  
           $(M_x\,M_y)=(15,\,5)$, $N=9$ whose size is 6.4 nm $\times$ 1.2 nm and hosts two edge states.}
  \label{Figure:magnetizations}
\end{figure}

\subsection{AFM-phase site-charge and -spin occupations}
The site-charge occupation can be split into bulk and edge contributions in the {\it edge-only doping} regime as 
follows:
\begin{eqnarray}
\label{eqn:densities}
n_{{\bf R}i}&=&n^B_{{\bf R}i}+n^E_{{\bf R}i}\\
n^B_{{\bf R}i}&=&\,\sum_{m \alpha}\,\frac{8\,f_{m,i}^2}{{\cal M}_x\,M_y\,\Lambda^\phi_{m\alpha}}\,
\left(\begin{matrix}(\phi_{m\alpha,i}^{\cal A})^2\\(\phi_{m\alpha,i}^{\cal B})^2\end{matrix}\right)\nonumber\\
n^E_{{\bf R}i}&=&\,\sum_{m}\, \frac{8\,f_{m,i}^2}{{\cal M}_x\,M_y\,\Lambda^\psi_m}\,\,
\left(\begin{matrix}(\psi_{m,i}^{\cal A})^2\\(\psi_{m,i}^{\cal B})^2\end{matrix}\right)\,n_m^E\nonumber
\end{eqnarray}
while the site spin densities are
\begin{eqnarray}
M^E_{{\bf R}i}&=&\sum_{m}\, \frac{8\,f_{m,i}^2}{{\cal M}_x\,M_y\,\Lambda^\psi_m}\,
\left(\begin{matrix}(-\psi_{m,i}^{\cal A})^2\\(\psi_{m,i}^{\cal B})^2\end{matrix}\right)\,P_m\nonumber
\end{eqnarray}
These occupations can be computed numerically. Some care must be
taken to handle numerical divergencies in the edge summations where hyperbolic sine functions appear.
We plot the edge magnetization of two rectangulenes in Figure \ref{Figure:magnetizations}.The first corresponds to
a wide but short one, that hosts a large number of edge states. The second one is a long 9-armchair GNR that
hosts two edge states only.

\subsection{There and back again: real-space tight-binding Hamiltonian}
\label{Subsection:tbmodel}
The mean-field Hubbard Hamiltonian can be written back in the site basis with the aid of 
equations (\ref{eqn:densities}) as follows
\begin{eqnarray}
\hat{H}_{MF}&=&\sum_{{\bf R}i \sigma}\sum_{a=A,B}\,\epsilon_{{\bf R} i \sigma}^a\hat{n}^a_{{\bf R} i \sigma} 
-t\,\sum_{< {\bf R} i \sigma, {\bf R'} i' \sigma' >}\left(\hat{a}^\dagger_{{\bf R} i\sigma} \hat{b}_{{\bf R'} i'\sigma'}+ c.c.\right)\nonumber\\
\epsilon_{{\bf R} i \sigma}^a&=&\left(\epsilon_0+U\,n^a_{R i \bar{\sigma}}\right)
\end{eqnarray}
This one-body Hamiltonian incorporates already both edge physics and electron correlations at the fully 
self-consistent mean field level. 
Writing the Hamiltonian back in the site basis in helpful because it can then be used to analyze further phenomena. Notice that quantum transport codes \cite{gollum,transiesta} take as input one-body tight-binding 
Hamiltonians. So by adding additional pieces to the Hamiltonian, more complex phenomena can be analysed.
Site / hopping disorder can be addressed by 
replacing $\epsilon_0$ / $t$ in the formula above with a random non-uniform distribution of on-site energies/hopping integrals 
$\epsilon_{{\bf R}_i \sigma}$ / $t_{{\bf R}_i \sigma,{\bf R'}_i' \sigma'}$. Similarly, a Peierls phase
$\phi_{{\bf R}_i,{\bf R}'_i}=\frac{\pi}{\phi_0}\,\int_{{\bf R}_i}^{{\bf R}'_i}\,\mathbf{A}\,\cdot\,d\mathbf{l}$ 
can be attached to the hopping integrals to investigate Hall physics, where $\phi_0$ is the flux quantum and 
$\mathbf{A}$ is the gauge vector associated to a uniform magnetic field $\mathbf{B}$ oriented along the $Z$-axis,
and solving numerically afterwards the rectangulene Hamiltonian. Coupling to a transverse gauge vector can
be included to analyze the optical response of the rectangulene to light. We note that even though the spin-dependent 
site occupations $n_{{\bf R} i \sigma}$ are no longer the exact solution of the Hamiltonian, the numerical solution
should deliver the qualitatively correct behavior/response of the rectangulene.

\section{Conclusions}
\label{Section:conclusions}
We have presented in this article a full analytical solution of the mean-field Hubbard model of non-chiral 
graphene rectangulenes of arbitrary length and width. A central result of the article has been the 
determination of the bulk, edge and cross Coulomb integrals of those rectangulenes, that are written here 
for the first time. This solution is not only an algebraic curiosity, but rather is a powerful and flexible
platform that may enable us to address a wide range of experimental issues in STM, transport, magnetic, Hall and 
optical phenomena of real-life graphene rectangulenes. It can also be used to address strong
electron correlations, by including GW or any other perturbative approach on top of it. 

\section*{Acknowledgements}
JF would like to thank Prof. Nazario Martin for confirming him that calling finite-length GNRs by the term 
{\it rectangulene} is chemically correct. This research has been funded by MCIN/AEI/10.13039/501100011033/ 
FEDER, UE via project PID2022-137078NB-100 and by Asturias FICYT under grant AYUD/2021/51185 with the support 
of FEDER funds.

\section*{Appendix}
We write in this appendix the explicit expressions for some coefficients appearing in the Coulomb integrals in section
\ref{Subsection:Coulombintegrals}.
We use the short-hands $k_\pm=(k_m^\alpha\pm k_{m'}^{\alpha'})/2$, $q_\pm=(q_m\pm q_{m'})/2$, $F_{\pm} = F(k_{m \alpha} \pm k_{m' \alpha'})$ and so forth.
\begin{widetext}
\begin{eqnarray}
C^{B,+}_{m\alpha,m'\alpha'}&=&2\,F^1_{m \alpha}+2\,F^1_{m' \alpha'}-F^1_{+}-F^1_{-} \nonumber\\
C^{B,-}_{m\alpha,m'\alpha'}&=&2\,F^2_{m \alpha}+2\,F^2_{m' \alpha'}-F^2_{+}-F^2_{-}\nonumber\\
C^{E,+}_{m, m'}&=&(\coth{({\cal M}_xq_m/2)}+\coth{({\cal M}_xq_{m'}/2)})\,G^1_{+}+(\coth{({\cal M}_xq_m/2)}-\coth{({\cal M}_xq_{m'}/2)})\,G^1_{-}-\nonumber\\
&&-\frac{2}{\sinh{({\cal M}_x q_m/2)}}\,G^1_{m'}-\frac{2}{\sinh{({\cal M}_x q_{m'}/2)}}\,G^1_{m}\nonumber\\
C^{E,-}_{m, m'}&=&(\coth{({\cal M}_xq_m/2)}\times\coth{({\cal M}_xq_{m'}/2)}+1)\,G^2_{+}+(\coth{({\cal M}_xq_m/2)}\times\coth{({\cal M}_xq_{m'}/2)}-1)\,G^2_{-}-\nonumber\\
&&-2\,\frac{\coth{({\cal M}_x q_m/2)}}{\sinh{({\cal M}_x q_{m'}/2)}}\,G^2_{m} -2\,\frac{\coth{({\cal M}_x q_{m'}/2)}}{\sinh{({\cal M}_x q_m/2)}}G^2_{m'}\nonumber\\
C^{BE,+}_{m\alpha,m'}&=&G^1_{m'}+\frac{1-F^1_{m \alpha}}{\sinh{({\cal M}_xq_{m'}/2)}}-\nonumber\\
&&-2\,\frac{\sinh{(q_{m'}/2)}\,\cos{({\cal M}_x\bar{k}_{m \alpha}/2)}\,\cos{(\bar{k}_{m \alpha}/2)}+
\cosh{(q_{m'}/2)}\,\coth{({\cal M}_x q_{m'}/2)}\,\sin{({\cal M}_x\bar{k}_{m \alpha}/2)}\,\sin{(\bar{k}_{m \alpha}/2)}}
{{\cal M}_x\,(\cosh{q_{m'}-\cos{\bar{k}_{m \alpha}})}}\nonumber\\
C^{BE,-}_{m\alpha,m'}&=&1+\coth{({\cal M}_xq_{m'}/2)}\,G^2_{m'}-\frac{F^2_{m \alpha}}{\sinh{({\cal M}_xq_{m'})}}-\nonumber\\
&&-2\,\frac{\coth{({\cal M}_xq_{m'}/2)}\,\cosh{(q_{m'}/2)}\,\cos{({\cal M}_x \bar{k}_{m \alpha}/2)}\,\cos{(\bar{k}_{m \alpha}/2)}+
\sinh{(q_{m'}/2)}\,\sin{({\cal M}_x\bar{k}_{m \alpha}/2)}\,\sin{(\bar{k}_{m \alpha}/2)}}
{\cosh{q_{m'}}+\cos{\bar{k}_{m \alpha}}} \nonumber
\end{eqnarray}
\end{widetext}

\bibliography{biblio.bib}

\end{document}